# Video Diffusion for Satellite-based High-Dynamical-Fidelity Precipitation (HiDFiP) Field Generation

Runze Li[a*], Yan Xia[b], Yongquan Qu[c,d], Dongwei Fu[e], Clement Guilloteau[a], Judy Hoffman[f], Stephan Mandt[f], Pierre Gentine[c,d], Efi Foufoula-Georgiou[a,b]

[a] Department of Civil and Environmental Engineering, University of California, Irvine, Irvine, CA 92697, USA
[b] Department of Earth System Science, University of California, Irvine, Irvine, CA 92697, USA
[c] Department of Earth and Environmental Engineering, Columbia University, New York, NY 10027, USA
[d] NSF Center for Learning the Earth with Artificial Intelligence and Physics (LEAP), Columbia University, New York, NY 10027, USA
[e] Space Science and Engineering Center, University of Wisconsin–Madison, Madison, WI 53706, USA
[f] Department of Computer Science, University of California, Irvine, Irvine, CA 92697, USA
[*] Corresponding author: Runze Li (runzl@umich.edu)

## Abstract

High spatiotemporal fidelity precipitation products, i.e., products that accurately capture the spatial organization, propagation, and lifecycle evolution of storms, are essential for advancing hydrometeorological research and operations at regional and global scales. Satellite precipitation products offer the only near-global precipitation observations, but they still fall far short of reproducing the spatiotemporal structure of ground-based reference measurements, due largely to dynamic distortions from the intrinsic inhomogeneity, intermittency, and indirectness of the satellite retrievals. Here we propose a video-diffusion framework for satellite-based High-Dynamical-Fidelity Precipitation (HiDFiP) field generation beyond the space–time coverage of ground-radar observations, and we use the radar-rich CONUS domain as a testbed to develop and evaluate the framework. The framework performs explicit spatiotemporal modeling with IMERG satellite precipitation as the primary source and leverages rich four-dimensional physical information of the storm environment extracted from ~40 atmospheric/land environmental fields from ERA5/ERA5-Land to compensate for the temporal information deficit inherent to satellite retrievals. We introduce an extensive suite of performance metrics to assess multifaceted dynamical fidelity of HiDFiP in both temporal-reconstruction and spatial-transfer settings, using MRMS ground-radar precipitation as a reference over CONUS. Relative to IMERG and image-wise diffusion baselines, HiDFiP accurately reproduces the storm space–time spectral characteristics; storm timing, location, and directional propagation; precipitation-event episodicity and temporal structure; precipitation-system morphology and spatial organization; and storm-track kinematics and lifecycle evolution. Transferability experiments indicate that HiDFiP generalizes reasonably well to an unseen region. This work advances a new video-diffusion paradigm for satellite-based, high-spatiotemporal-fidelity precipitation field generation and provides both an algorithmic and diagnostic foundation for developing long-term, global radar-grade precipitation records.

## 1. Introduction

Precipitation is a fundamental yet complex component of Earth’s water and energy cycles. High-quality precipitation datasets therefore underpin a wide spectrum of hydrology, meteorology, and ecology studies (Jiang et al., 2023; Qin et al., 2025; Shiogama et al., 2022). Given the multiscale structure and highly skewed, intermittent statistics of precipitation, applications such as flood forecasting and convective-scale diagnostics require more than accurate mean values or accumulations (Hettiarachchi et al., 2018; Rajagopal et al., 2023); rather, they demand faithful storm organization, timing/location, and lifecycle evolution, collectively referred to here as “spatiotemporal fidelity” (O'Brien et al., 2016). Meeting this standard, in turn, requires precipitation datasets that are simultaneously accurate and spatiotemporally continuous. Gauge-adjusted ground-radar Quantitative Precipitation Estimation (QPE) arguably comes closest to this ideal and is therefore often treated as a near-ground-truth reference (Ayat et al., 2021; Li et al., 2023a, b), yet its availability is severely limited by uneven infrastructure deployment and data-access policies (Saltikoff et al., 2019). In contrast, spatiotemporally seamless satellite precipitation products, although commonly treated as observational references in numerical model evaluation and climate projection constraints (Ahn et al., 2023; Shiogama et al., 2025), carry systematic uncertainties rooted in indirect retrieval physics and intermittent sampling from space (Kidd et al., 2021; Xiong et al., 2025). Together, these limitations impose a persistent quality–coverage trade-off in precipitation records, motivating efforts to reconcile radar fidelity with satellite coverage, an objective for which state-of-the-art deep learning architectures offer unprecedented potential (Dai and Ushijima-Mwesigwa, 2025; Guilloteau et al., 2025; Wang et al., 2021).

Among modern deep-learning paradigms, generative diffusion models, which have revolutionized image/video synthesis in recent years by effectively learning complex, high-dimensional distributions with high perceptual realism (Ho et al., 2020; Ho et al., 2022b; Yang et al., 2023), have also demonstrated exceptional utility in atmospheric science and remote sensing (Dai et al., 2025; Pathak et al., 2026; Price et al., 2025). Unlike deterministic regression models that often yield blurred predictions, they excel at capturing the fine-scale structure and extremes of precipitation, enabling probabilistic ensemble generation that remains statistically consistent with ground truth, ensuring high-fidelity outputs (Liu et al., 2025a; Wang et al., 2024a; Yu et al., 2024). Yet, compared with the rapidly growing literature on diffusion-based probabilistic precipitation prediction (e.g., nowcasting/forecasting) and model-output post-processing (e.g., downscaling/bias-correction) tasks (Gao et al., 2023; Srivastava et al., 2024; Xu et al., 2026), efforts specifically targeting high-fidelity satellite-based precipitation product development remain comparatively limited (Dai and Ushijima-Mwesigwa, 2025; Guilloteau et al., 2025), even though their fidelity constitutes the foundational basis for the many aforementioned downstream tasks in which they are often treated as pseudo-ground-truth training targets (Harder et al., 2026; Stock et al., 2024; Yuval et al., 2026).

Distinct from numerical simulations, remotely sensed precipitation products exhibit unique error modes, among which temporal inconsistency is particularly consequential (Li et al., 2025a; Li et al., 2025b; Li et al., 2018). Passive Microwave (PMW) sensors, the primary observational basis for satellite precipitation retrieval, are largely confined to (near-)polar orbits due to the antenna aperture constraints at geostationary altitudes (Kidd et al., 2021, 2022), which severely limit native revisit frequency and directly bound the effective temporal resolution of any downstream merged products (Guilloteau et al., 2021, 2022). However, even with hypothetically uninterrupted PMW sampling, precipitation retrieval remains more weakly constrained by observed radiances than many other geophysical variables (Guilloteau and Foufoula-Georgiou, 2020; Guilloteau et al., 2018), hindering the resolution of rapidly evolving cloud microphysics and introducing storm-stage-dependent errors well documented in prior studies (Guilloteau and Foufoula-Georgiou, 2024; Li et al., 2025b; Li et al., 2021). Under such weak temporal constraints and given precipitation's intrinsically high variability, employing image-based diffusion models entails stochastic sampling from a relatively broad conditional distribution over plausible precipitation fields. Without explicit inter-frame coupling, independent samples at adjacent time steps may therefore diverge substantially, producing "flickering" artifacts and temporal inconsistencies that can even exceed those of simpler deterministic baselines such as UNet (Jeong and Ye, 2024; Wu et al., 2023).

To address these challenges, here we emphasize the necessity of explicit temporal modeling in satellite-based precipitation product development and propose a video-diffusion framework for High-Dynamical-Fidelity Precipitation (HiDFiP) field generation. This framework enforces spatiotemporal statistical consistency with ground-radar observations, with the goal of reconstructing high-fidelity precipitation beyond radar's limited spatiotemporal coverage. We demonstrate the framework over CONUS, which serves as a radar-rich testbed for both temporal extrapolation and spatial transfer. To compensate for the unavoidable temporal information deficit in satellite products, and building on our previous finding that the temporal structure of environmental variables provides physically interpretable constraints on precipitation retrievals (Li et al., 2025a), we condition the model on ~40 four-dimensional (longitude–latitude–level–time) atmospheric and land predictors alongside the raw satellite precipitation product, and conduct grouped ablation experiments to systematically quantify their contributions. Built on a 3D UNet video diffusion backbone, our framework initializes from a temporally structured noise prior and proceeds through a two-stage refinement under a composite objective that incorporates a Wasserstein Distance penalty for enhanced distributional alignment. Finally, moving beyond conventional verification, we develop a comprehensive hierarchical diagnostic suite to assess the multifaceted dynamical fidelity of HiDFiP, e.g., wavenumber–frequency spectral analysis and process-oriented diagnostics, including precipitation event-, system-, and track-based metrics, to characterize the multiscale storm propagation and lifecycle evolution critical to hydrometeorological applications.

## 2. Data

### 2.1. IMERG

Satellite precipitation estimates are taken from the NASA Global Precipitation Measurement (GPM) Integrated Multi-satellitE Retrievals for GPM (IMERG) V07B Final Run (Huffman et al., 2023). IMERG provides global 0.1°×0.1°, 30-min surface precipitation rates by merging PMW retrievals from the GPM constellation with Infrared (IR)-based estimates and applying monthly gauge adjustment based on Global Precipitation Climatology Centre (GPCC) analyses. IMERG provides a Quality Index (QI; 0–1, higher is better), designed as a composite-correlation proxy of expected retrieval skill that varies with the evolving mix of PMW- and IR-based inputs (Huffman, 2019).

### 2.2. ERA5/ERA5-Land

The environmental predictors used here include atmospheric variables from the European Centre for Medium-Range Weather Forecasts (ECMWF) Reanalysis Version 5 (ERA5), land-surface variables and topography (represented by surface geopotential) from ERA5-Land (Hersbach et al., 2023a, b; Muñoz Sabater, 2019). ERA5 is produced with 4D-Var data assimilation and is provided hourly on a regular 0.25° grid. ERA5-Land further builds on ERA5 by replaying the land component offline at higher spatial resolution (0.1°), driven by ERA5 atmospheric forcing to better represent land hydrology and soil processes. We extract 37 predictor fields, organized into five physical categories: topography, dynamics (e.g., vertical velocity, geopotential height), thermodynamics (e.g., temperature, convective available potential energy), moisture (e.g., specific humidity, moisture divergence), and cloud properties (e.g., cloud water content, total cloud cover) (see Appendix A for details). Note that, given the well-documented biases in ERA5-family precipitation and the inclusion of IMERG precipitation as input (Wang et al., 2026; Wu et al., 2024), we exclude ERA5/ERA5-Land precipitation from the inputs to avoid introducing a potentially redundant and biased precipitation proxy that may encourage shortcut learning. All predictors are bilinearly remapped to the 0.1° grid and linearly interpolated from hourly to 30-min resolution.

### 2.3. MRMS

Ground-truth precipitation is provided by the Multi-Radar Multi-Sensor (MRMS) V12 QPE, which integrates dual-polarization observations from the WSR-88D radar network, supplemented with model/climatological information, and bias-corrected with gauge data (Zhang et al., 2016; Zhang et al., 2020). MRMS also provides a Radar Quality Index (RQI; 0–1, higher is better) that combines beam-blockage and beam-height/range effects, which can be roughly regarded as a proxy for effective radar coverage and range (Zhang et al., 2012). Since the highest-quality multi-sensor QPE is available only at hourly resolution, whereas radar-only QPE is provided at 15-min intervals, we derive 30-min MRMS precipitation by computing an hourly multiplicative factor (multi-sensor QPE divided by hourly-aggregated radar-only QPE) and applying it to 30-min radar-only accumulations. This procedure follows established practice in NASA's GV-MRMS dataset and prior studies (Guilloteau et al., 2025; Kirstetter et al., 2012). The native 0.01° MRMS grids are aggregated to 0.1° to match the target resolution.

# 3. Method

## 3.1. Overview

Figure 1 visualizes the study overview, including the problem framing (panel A), study setup (panel B), technical motivation (panel C), proposed framework (panel D), and evaluation suite (panel E).

**Problem Framing** Panel A frames the fundamental quality–coverage trade-off across precipitation-related products, motivating our effort to bridge the gap. Datasets are conceptually placed in a quality–coverage space: ground-radar QPE anchors the high-quality/low-coverage corner, whereas satellite precipitation products and reanalysis environmental fields lie toward low-quality/high-coverage areas. The deep generative model pursued here aims to extend the envelope toward the high-quality/high-coverage regime, with a long-term vision of radar-grade fidelity at satellite-scale spatiotemporal extent. As a step toward this vision, we use CONUS as a testbed for algorithm development and reconstruction/transferability experiments.

## A. Problem Framing: High-fidelity Precipitation Beyond Radar Spatiotemporal Coverage

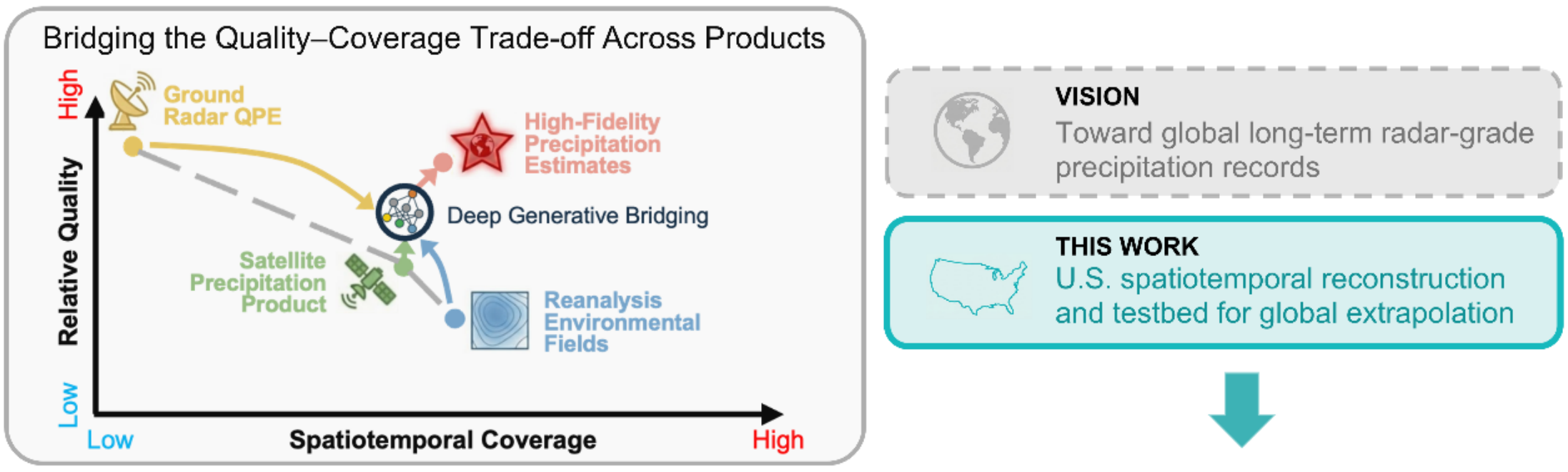

## B. Study Setup: CONUS Testbed for Reconstruction and Transfer

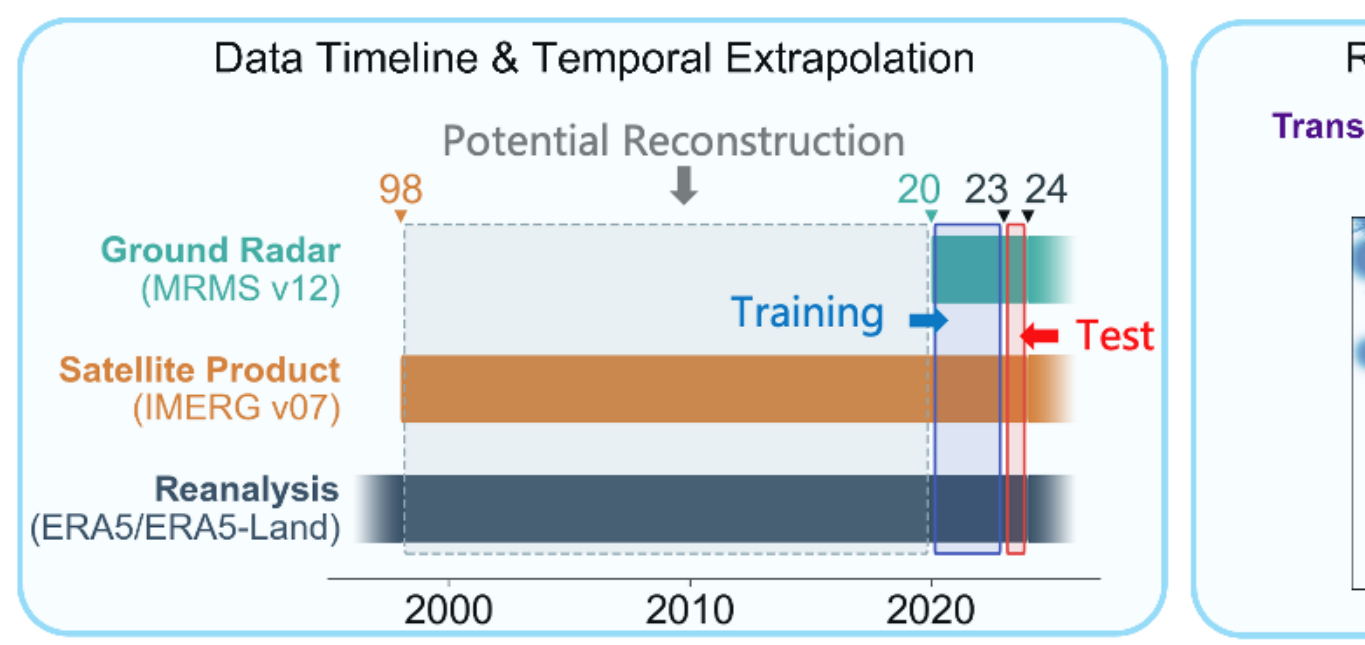

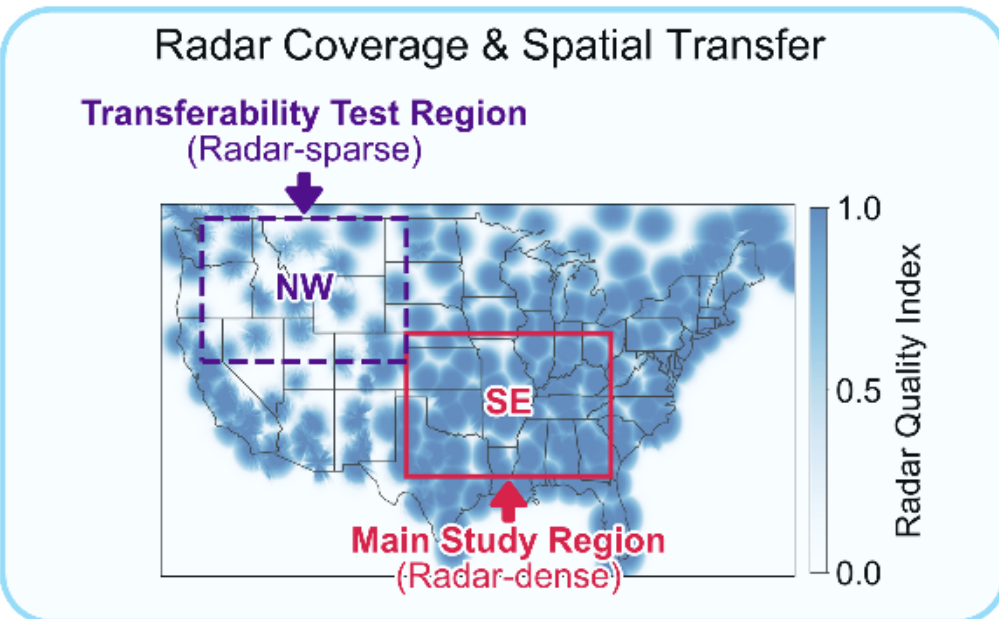

## C. Technical Motivation: Necessity of Explicit Temporal Modeling

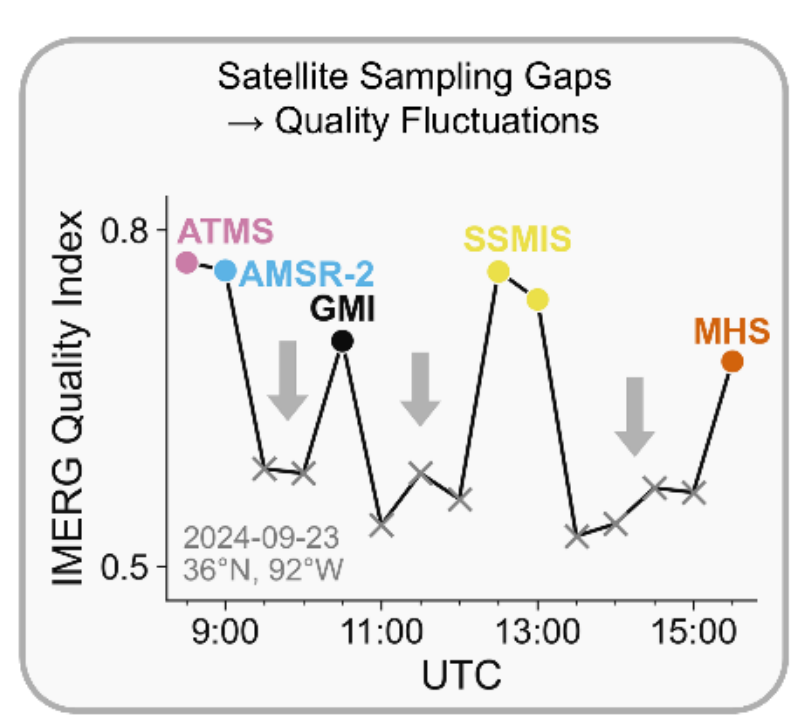

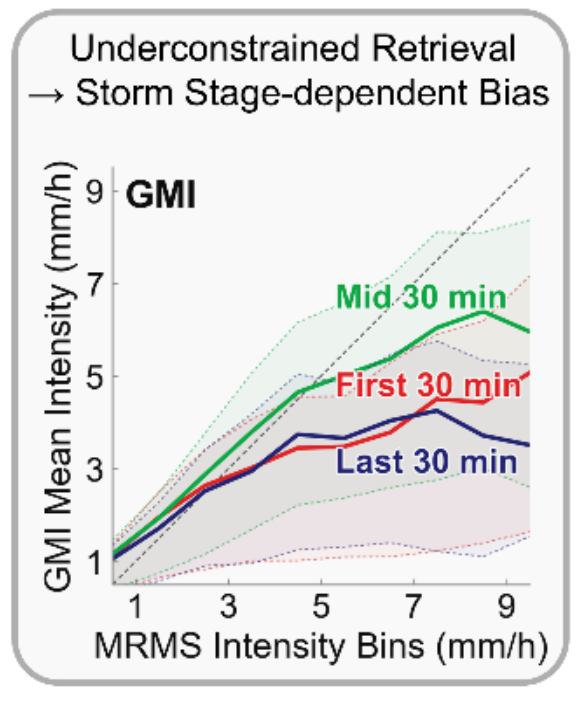

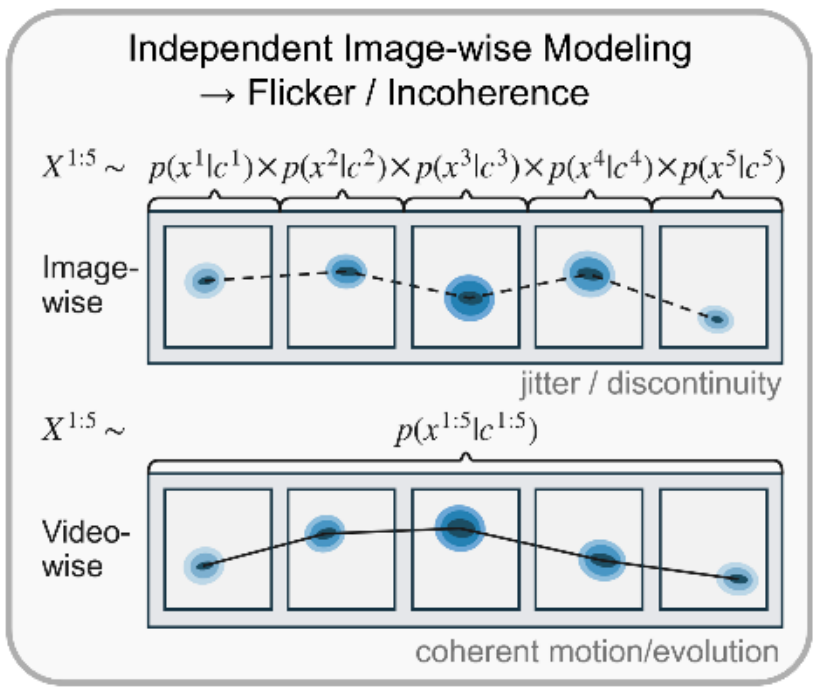

## D. Proposed Framework: Video Diffusion for High Spatiotemporal Fidelity Precipitation

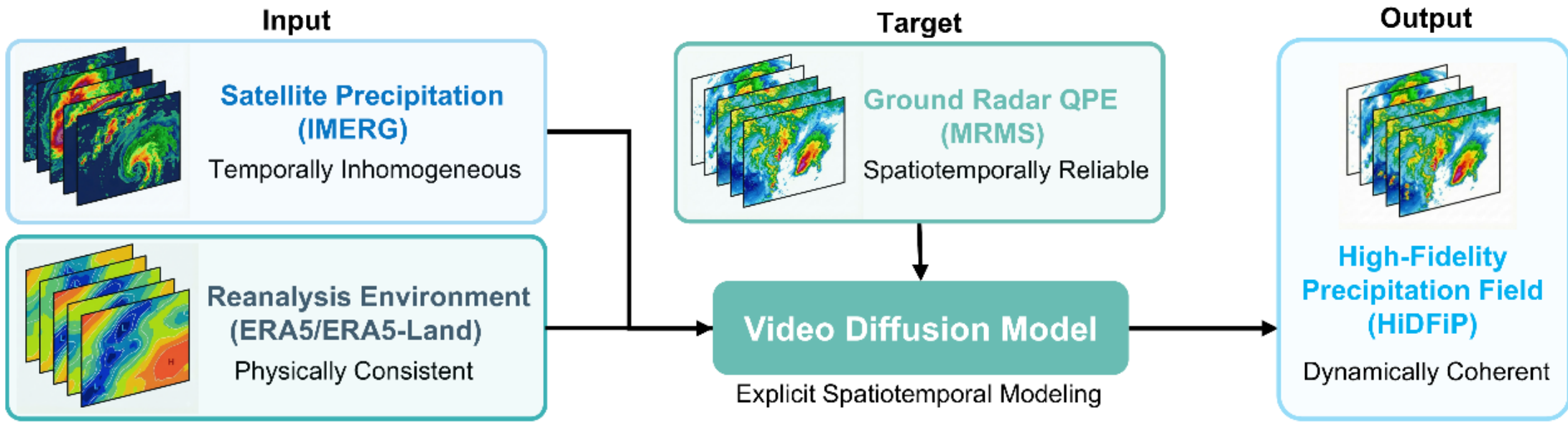

## E. Evaluation Suite: Hierarchical, Multifaceted Diagnostics

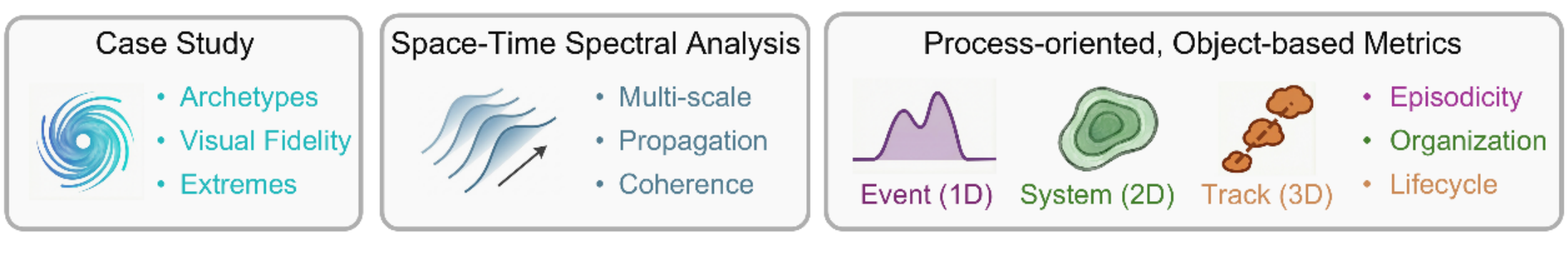

**Figure 1** Study Overview. **(A) Problem framing**: Bridging the quality–coverage trade-off toward long-term radar-grade precipitation estimates. **(B) Study setup**: Temporal extrapolation and spatial transfer experiments over CONUS. **(C) Technical motivation**: Satellite sampling gaps, underconstrained retrieval, and independent image-wise modeling. **(D) Proposed framework**: Video-diffusion pipeline for High-Dynamical-Fidelity Precipitation (HiDFiP) field generation. **(E) Evaluation suite**: Hierarchical diagnostics spanning case studies, spectral analysis, and process-oriented metrics.

**Study Setup** Panel B specifies this CONUS testbed by delineating MRMS spatiotemporal coverage constraints and the corresponding experimental setup. Temporally (left panel B), MRMS V12 has a comparatively short record (available since October 2020) relative to IMERG V07 (since 1998) and ERA5/ERA5-Land (since 1940/1950), which naturally motivates the need for temporal reconstruction. Notably, while earlier MRMS versions extend further back in time, NOAA implemented the most substantial MRMS QPE upgrade since the system's initial operations with V12 in Oct 2020 (Gerard et al., 2021), without retrospectively processing the pre-V12 archive. We therefore define MRMS availability in this study by the V12 record used here. To evaluate temporal generalization, we train the model on Oct 2020–Sep 2023 and test it on Oct 2023–Sep 2024. Spatially (right panel B), the MRMS RQI map presented here highlights uneven radar coverage across CONUS, underscoring the significance of spatial reconstruction. Accordingly, we designate the radar-dense Southeast (SE) CONUS (31°–41°N, 102°–82°W) as the primary study region, where reference data are highly reliable, and the radar-sparse Northwest (NW) CONUS (39°–49°N, 122°–102°W) as the spatial transferability test region.

**Technical Motivation** Panel C isolates the technical motivation for explicit temporal modeling for satellite precipitation. The left panel shows pronounced fluctuations in IMERG QI within a representative time window, exposing the fundamental limitation—temporally inhomogeneous retrieval quality—in high-resolution multi-sensor satellite precipitation products, driven by intermittent PMW sampling. Periods coinciding with PMW overpasses (*ATMS: Advanced Technology Microwave Sounder, AMSR-2: Advanced Microwave Scanning Radiometer 2, GMI: GPM Microwave Imager, SSMIS: Special Sensor Microwave Imager/Sounder, and MHS: Microwave Humidity Sounder*; marked by circles) exhibit markedly higher quality than intervals without PMW observations (marked by crosses). This behavior constitutes a key reason why satellite precipitation products particularly require explicit temporal modeling to propagate information across observational gaps. The middle panel, redrawn and adapted from our previous study (Li et al., 2025b), shows distinct intensity correspondences between ground-based MRMS observations and single-sensor GMI retrievals across different storm stages (represented by color-coded solid lines for the first, middle, and last 30 min of precipitation events; shaded bands indicate the 25th–75th percentile range). This stage dependence indicates that, even under idealized, temporally seamless single-sensor coverage, retrieval errors would remain nonstationary because the intrinsically underconstrained mapping from PMW signals to surface precipitation cannot adequately respond to the evolving cloud microphysics throughout storm lifecycles. Temporal context must therefore be explicitly incorporated to regularize this intrinsically ambiguous retrieval relationship over time. The right panel conceptually contrasts image-wise and video-wise generation in representing storm evolution. Image-wise generation samples each frame independently, i.e., $\boldsymbol{X}^{1:5} \sim \prod_{t=1}^{5} p(\boldsymbol{x}^t \mid \boldsymbol{c}^t)$ (where $\boldsymbol{X}^{1:5}$ is the generated 5-frame sequence used here for illustration, and $\boldsymbol{x}^t$ and $\boldsymbol{c}^t$ are the frame and condition at time $t$, respectively), ignoring inter-frame coupling and, owing to generative sampling stochasticity, nearly inevitably yielding realizations with incoherent dynamics. This issue is particularly acute in satellite-based precipitation generation because the observational gaps and retrieval ambiguity illustrated above, compounded by precipitation's intrinsic high variability, broaden the range of plausible frame-wise realizations and thereby exacerbate frame-to-frame divergence and flickering. In contrast, video-wise generation explicitly models temporal dependencies and consistently draws joint sequence samples adhering to realistic dynamics, i.e., $\boldsymbol{X}^{1:5} \sim p(\boldsymbol{x}^{1:5} \mid \boldsymbol{c}^{1:5})$ (where $\boldsymbol{x}^{1:5}$ and $\boldsymbol{c}^{1:5}$ denote the joint frame and condition sequences), thereby enforcing physically plausible temporal evolution that preserves storm motion and lifecycle.

**Proposed Framework** Panel D presents the schematic overview of the proposed video diffusion framework. The model ingests temporally inhomogeneous satellite precipitation (IMERG) and physically consistent reanalysis environmental fields (ERA5/ERA5-Land), learns to reconstruct spatiotemporally reliable ground-radar QPE (MRMS) through a video diffusion model with explicit spatiotemporal modeling, and generates high-dynamical-fidelity precipitation fields, HiDFiP, that preserve dynamical coherence in both space and time.

**Evaluation Suite** Panel E illustrates the dedicated hierarchical evaluation framework we designed to comprehensively interrogate the multifaceted spatiotemporal dynamics of HiDFiP and benchmark products. The first tier provides intuitive case studies highlighting archetypal patterns, visual fidelity, and representative extreme events. The second tier employs nonparametric space–time spectral analysis to quantify multiscale dynamics, storm propagation, and spatiotemporal coherence. The third tier introduces parametric object-based diagnostics across three structural levels: precipitation-events (1D), -systems (2D), and -tracks (3D), enabling systematic assessment of storm episodicity, organization, and lifecycle evolution. This novel multilevel evaluation framework enables a comprehensive and physically meaningful interpretation of model dynamical fidelity, which is critical for hydrometeorological applications. See Section 3.6 for details.

### 3.2. Model Workflow

Figure 2 provides an overview of the video-diffusion framework for HiDFiP generation, which adopts a two-stage mean–residual design shown to offer advantages over direct diffusion modeling in precipitation-related tasks (Guilloteau et al., 2025;

Srivastava et al., 2024). The input comprises the channel-wise concatenation of video-formatted IMERG precipitation and ERA5/ERA5-Land atmospheric/land-surface states and is denoted by $\boldsymbol{x} \in \mathbb{R}^{B \times C \times F \times H \times W}$, where $B$ is the batch size, $C$ is the number of input channels, $F$ is the number of temporal frames, and $H \times W$ is the spatial grid size. In this study, each input clip comprises $F = 20$ consecutive half-hourly frames over the full $H \times W = 100 \times 200$ grid of the corresponding SE or NW study domain. The input is first passed through a deterministic 3D Regression UNet (R-UNet) to produce a coarse precipitation estimate that can be empirically interpreted as the conditional mean signal: $\overline{\boldsymbol{y}}_\theta = \mu_\theta(\boldsymbol{x})$, where $\mu_\theta(\cdot)$ denotes the network parameterized by $\theta$. The input together with the estimated mean $\overline{\boldsymbol{y}}_\theta$ then jointly conditions a video diffusion module equipped with a 3D Denoising UNet (D-UNet) that models the residual $\boldsymbol{r}_0 = \boldsymbol{y} - \overline{\boldsymbol{y}}_\theta$, where $\boldsymbol{y} \in \mathbb{R}^{B \times 1 \times F \times H \times W}$ is the ground-truth MRMS target. Specifically, the diffusion model learns the conditional residual distribution $p_\phi(\boldsymbol{r}_0 \,|\, \boldsymbol{x}, \overline{\boldsymbol{y}}_\theta)$ parameterized by $\phi$ to synthesize fine-scale stochastic precipitation structures. The cascaded R-UNet and D-UNet are jointly optimized end-to-end. Sampling starts from a structured noise prior $\boldsymbol{r}_T$, detailed in the following section, and proceeds through $K$ denoising steps using Denoising Diffusion Implicit Models (DDIM) to obtain $\hat{\boldsymbol{r}}_0$ (Song et al., 2021). Drawing $M$ independent residual realizations $\{\hat{\boldsymbol{r}}_0^{(i)}\}_{i=1}^{M}$, the final ensemble predictions are obtained as: $\hat{\boldsymbol{y}}^{(i)} = \overline{\boldsymbol{y}}_\theta + \hat{\boldsymbol{r}}_0^{(i)}$, $i = 1, 2, \dots, M$. The detailed internal model architecture is provided in Appendix B.

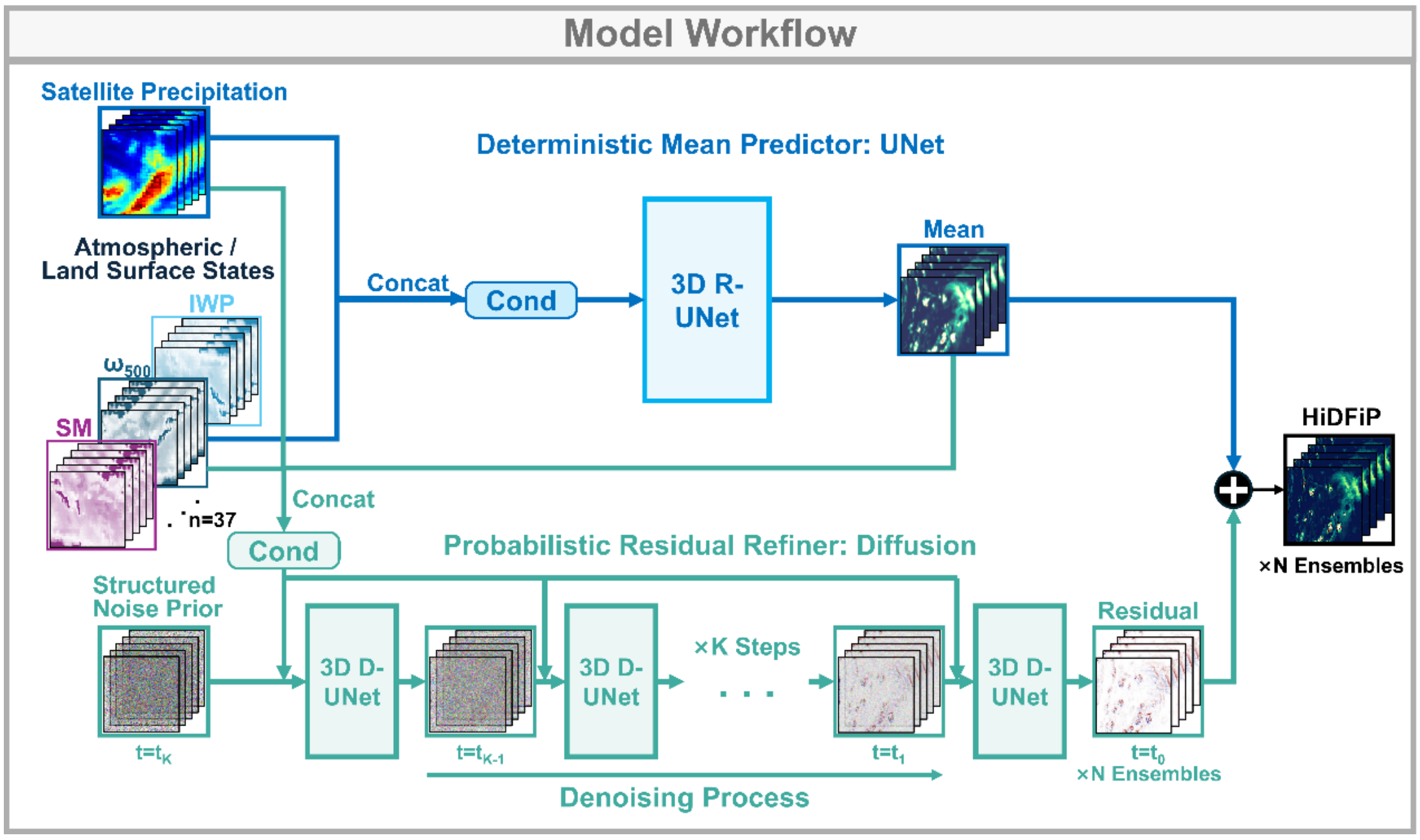

**Figure 2** Model Workflow. The framework adopts a cascaded two-stage design: A 3D Regression UNet (R-UNet) first produces a deterministic mean precipitation estimate from video-formatted IMERG/ERA5/ERA5-Land inputs. A video-diffusion module with a 3D Denoising UNet (D-UNet) then predicts the residual relative to MRMS from a structured noise prior, conditioned on the same inputs alongside the estimated mean,

generating an ensemble of residual predictions that are added back to the estimated mean to form the final video-formatted HiDFiP ensembles.

### 3.3. Video Diffusion Formulation and Training Objective

Diffusion models learn data distributions via an iterative noising-denoising process. Unlike standard video diffusion formulations that inject independent Gaussian noise across frames (Ho et al., 2022b), we employ a structured noise prior that introduces explicit inter-frame correlation to further enhance temporal consistency (Ge et al., 2023). Specifically, for each frame $i$, the noise is constructed as a variance-preserving mixture of a frame-shared component $\boldsymbol{\epsilon}_{\text{shared}}$ and a frame-specific component $\boldsymbol{\epsilon}_{\text{ind}}^{i}$,

$$\boldsymbol{\epsilon}^{i} = \boldsymbol{\epsilon}_{\text{shared}} + \boldsymbol{\epsilon}_{\text{ind}}^{i}. \tag{1}$$

This construction preserves the per-frame marginal covariance while inducing a constant cross-frame covariance, thereby embedding temporal coupling directly in the perturbation process without requiring additional conditioning pathways or learnable parameters. Stacking the frame-wise noise maps yields $\boldsymbol{\epsilon} \sim \mathcal{N}(\mathbf{0}, \boldsymbol{\Sigma})$, where $\boldsymbol{\Sigma}$ denotes the resulting structured spatiotemporal covariance matrix. This temporal inductive bias is intended to promote smoothly evolving trajectories consistent with the physical continuity of precipitation fields, particularly in the limited-sample video setting considered here.

We use a conditional video diffusion framework to model the joint residual distribution $p_{\phi}(\boldsymbol{r}_0 \mid \boldsymbol{x}, \overline{\boldsymbol{y}}_{\theta})$ over precipitation sequences (with $p_{\phi,\theta}(\boldsymbol{y} \mid \boldsymbol{x})$ recovered via $\boldsymbol{y} = \overline{\boldsymbol{y}}_{\theta} + \boldsymbol{r}_0$). Training follows the Denoising Diffusion Probabilistic Model (DDPM)-style forward noising process (Ho et al., 2020), whereas inference uses a Denoising Diffusion Implicit Model (DDIM)-style reverse denoising trajectory for accelerated inference (Song et al., 2021). Specifically, we retain the scalar noise schedule $\{\beta_t\}_{t=1}^{T}$, with $\alpha_t = 1 - \beta_t$, $\bar{\alpha}_t = \prod_{s=1}^{t} \alpha_s$, and $\bar{\alpha}_0 = 1$, but replace the identity covariance $\mathbf{I}$ in standard DDPM derivations with the structured noise covariance $\boldsymbol{\Sigma}$ defined above. The resulting forward marginal for the noised residual state $\boldsymbol{r}_t$ at timestep $t$ is

$$q(\boldsymbol{r}_t \mid \boldsymbol{r}_0) = \mathcal{N}\big(\boldsymbol{r}_t; \sqrt{\bar{\alpha}_t}\boldsymbol{r}_0, (1 - \bar{\alpha}_t)\boldsymbol{\Sigma}\big). \tag{2}$$

During reverse denoising, residual generation starts from $\boldsymbol{r}_T \sim \mathcal{N}(\mathbf{0}, \boldsymbol{\Sigma})$ and proceeds through $K$ DDIM sampling steps along the selected timestep sequence $T = t_K > \cdots > t_0 = 0$, where $K < T$. At step $k$, the D-UNet conditions on the current noised residual $\boldsymbol{r}_{t_k}$, timestep $t_k$, and conditioning variables $(\boldsymbol{x}, \overline{\boldsymbol{y}}_{\theta})$ to parameterize the transition to $t_{k-1}$:

$$\boldsymbol{r}_{t_{k-1}} \sim p_{\phi}\big(\boldsymbol{r}_{t_{k-1}} \mid \boldsymbol{r}_{t_k}, \boldsymbol{x}, \overline{\boldsymbol{y}}_{\theta}\big), \quad k = K, \ldots, 1. \tag{3}$$

Iterating these updates yields the final residual sample $\hat{\boldsymbol{r}}_0$, which is then added back to the deterministic estimate $\bar{\boldsymbol{y}}_{\theta}$ to yield a generated precipitation field. Ensemble predictions are obtained by repeating the DDIM trajectory with independent noise realizations.

We adopt the $v$-parametrization as the denoising objective, denoted by $\mathcal{L}_v$, which provides a stable denoising target across diffusion times. In practice, the diffusion loss alone tends to underrepresent heavy-rain tails. We therefore augment the diffusion objective with a Wasserstein Distance regularization term, denoted by $\mathcal{L}_{\text{SWD}}$, which enforces distributional alignment between predictions and targets, implemented efficiently through the Sliced Wasserstein-1 Distance (SWD) (Liu et al., 2025b). The final objective is

$$\mathcal{L}(\theta,\phi) = (1-\omega)\mathcal{L}_v(\theta,\phi) + \omega\mathcal{L}_{\text{SWD}}(\theta,\phi),\ \omega \in [0,1], \tag{4}$$

where $\omega$ is the trade-off weight between the two loss terms, which is set to 0.02 based on validation experiments. Additional formulation and loss function details are provided in Appendix C. Model implementation details are provided in Supplementary Method S1.

### 3.4. Baseline and Ablation Design

**Baselines** Given the limited amount of prior work applying diffusion models to satellite-based precipitation estimation, and the fact that the few related studies mostly rely on standard image-based diffusion backbones rather than task-specific architectures (Dai and Ushijima-Mwesigwa, 2025), we implemented several canonical baselines in-house for comparison, with particular emphasis on quantifying the added value of introducing the video diffusion model for modeling spatiotemporally coherent precipitation dynamics: (1) Attention UNet: A deterministic UNet with attention modules (Oktay et al., 2018); (2) DDPM: A standard Denoising Diffusion Probabilistic Model (DDPM) with UNet denoiser (Ho et al., 2020); (3) ResDDPM: Residual DDPM, a classic two-stage UNet regressor + DDPM residual refiner for precipitation tasks (Guilloteau et al., 2025).

**Ablation Studies** Rather than emphasizing incremental gains from architectural modules, our ablation experiments isolate the contributions of different categories of 4D environmental predictors to clarify how each group provides complementary, physically interpretable spatiotemporal constraints beyond the information contained in satellite precipitation. We also remove IMERG from the inputs to assess the predictive skill achievable from environmental predictors alone. Specifically, we construct seven grouped ablation settings by removing: (1) satellite precipitation, (2) all environmental variables, and one environmental predictor family at a time: (3) topography, (4) dynamics, (5) thermodynamics, (6) moisture, and (7) cloud properties.

### 3.5. Spatial Transferability Test

In this section, we do not attempt a dedicated transfer-learning strategy, which is left for future work. Instead, we ask whether the diffusion model trained over the SE region, which jointly exploits spatiotemporal structure and environmental predictors (Figure 1), can retain useful skill when applied directly to the unseen NW region.

Because MRMS contains extensive low-quality coverage in this region, evaluation is restricted to high-RQI pixels (RQI > 0.8). Over this subset, we further apply a simple empirical quantile mapping in a five-fold cross-validation manner to obtain HiDFiP-QM, allowing us to assess whether a modest distributional correction can more clearly reveal the model's ability to recover storm dynamics under direct regional transfer. Given the sparsity of high-RQI coverage over NW and the random sample selection inherent in the cross-validation, the resulting training points are spatially dispersed. This experiment therefore partially simulates the effect of gauge-based correction in regions entirely lacking radar coverage.

### 3.6. Evaluation Framework

The dedicated hierarchical evaluation suite for multifaceted dynamical fidelity of the generated precipitation fields spans case studies, distributional/ensemble metrics, spectral analyses, process-oriented diagnostics at the precipitation-event, -system, and -track levels, and compact scalar metrics. Except for the ensemble-based metrics, namely Rank Histogram, Attribute Diagram, Sharpness Histogram, CRPS, and twCRPS, all metrics are first computed separately for each ensemble member and only then averaged across members, thereby avoiding the smoothing of fine-scale details that would result from premature ensemble averaging.

**Case Studies** Case studies provide an initial visual assessment of storm evolution and propagation before population-level statistics. Representative storm cases are examined over selected time windows using half-hourly precipitation snapshots and longitude–time/latitude–time Hovmöller diagrams, which average precipitation along the orthogonal spatial dimension to summarize zonal and meridional propagation through time.

**Distributional/Ensemble Metrics** Distributional and ensemble metrics jointly evaluate the statistical realism and probabilistic reliability of the generated precipitation fields. These include (1) Probability Density Function (PDF), which assesses agreement with MRMS in the marginal intensity distribution, especially in the heavy-rain tail; (2) Rank Histogram, which tests ensemble dispersion by examining the exchangeability of MRMS observations with ensemble members; (3) Attribute Diagram, which quantifies binary occurrence-probability calibration at the rainy threshold; (4) Sharpness Histogram, which indicates whether predicted probabilities are informative rather than concentrated near climatology. Details of these metrics are provided in Appendix D.1.

**Spectral Diagnostics** Spectral diagnostics offer a nonparametric, scale-resolved characterization of precipitation dynamics by projecting spatiotemporal precipitation cubes into joint frequency–wavenumber space via a three-dimensional Fourier transform, which enables a direct assessment of scale-dependent variance, propagation behavior, and spatiotemporal alignment of precipitation fields (Guilloteau et al., 2021). To make this spectral representation physically interpretable, the 3D spectra are first marginalized into univariate spectra along temporal, zonal, and meridional dimensions to isolate dimension-specific spectral variability. From the marginalized auto- and cross-spectra, we derive three complementary diagnostics: (1) Power Spectral Density (PSD) for variance distribution across temporal and spatial scales, (2) Coherence for spectral agreement between evaluated and reference precipitation fields, and (3) Phase Difference for cross-spectral phase offsets that indicate temporal lags or spatial shifts. The spectra are further marginalized into bivariate joint spectra, including wavenumber–frequency spectra for storm propagation and two-dimensional wavenumber spectra for storm morphology and anisotropy. Full spectral estimation procedures are provided in Appendix D.2.

Additionally, to provide a compact summary of multiscale propagation, we derive band-limited net phase-velocity vectors from the 3D spectra for each prescribed wavelength band. Within each wavelength band, component phase-velocity vectors are aggregated using PSD-weighted averaging to obtain a single summary vector representing the dominant propagation direction and net phase speed at that scale.

**Process-oriented Diagnostics** Process-oriented diagnostics use feature extraction approaches to identify precipitation objects explicitly tied to storm processes and evaluate their physically meaningful attributes, providing both a stringent test of model fidelity and a practical benchmark for hydrometeorological applications. We assess the process-oriented performance of HiDFiP from three complementary perspectives: precipitation event (one-dimensional, temporally continuous rainy series at fixed location), precipitation system (two-dimensional, spatially continuous rainy area at fixed time), and precipitation track (three-dimensional, spatiotemporally continuous rainy objects evolving across space and time). A common rain threshold of $\tau = 0.1$ mm/h is used throughout. Definitions and notation for the event-, system-, and track-level scalar metrics are provided in Appendices D.3.1.–D.3.3., respectively.

- **Precipitation Event** A precipitation event is defined as a temporally continuous rainy sequence at a given location, bounded by dry periods (Li et al., 2023a). Events intersecting the temporal boundaries of a clip are excluded from the analysis. Eight representative properties are characterized to describe precipitation-event episodicity and temporal organization: (1) Count, (2) Depth, (3) Duration, (4) Intensity, (5) Peak/Mean Ratio, (6) Peak Timing, (7) Short Extremes, and (8) Intermittency.
- **Precipitation System** A precipitation system is defined as a spatially contiguous rainy region within a given frame, identified using 8-neighbor connectivity (Li et al., 2022). Systems intersecting the spatial boundaries of a clip are excluded from the analysis. Additionally, we fit an equivalent ellipse to each system using the second-order central moments of its binary mask (Zhang et al., 2023a). Eight representative metrics are characterized to describe precipitation-system morphology and spatial organization: (1) Scale, (2) Orientation, (3) Aspect Ratio, (4) Regularity, (5) Asymmetry, (6) Concentration, (7) Radial Decay, and (8) Core Area.
- **Precipitation Track** A precipitation track is defined as a spatiotemporally coherent trajectory of a rainy object, constructed by Lagrangian object-based tracking across successive frames (Guilloteau and Foufoula-Georgiou, 2024). Tracking is performed using the Tracking and Object-Based Analysis of Clouds (tobac) framework (Heikenfeld et al., 2019). In each frame, rainy regions are identified and separated into individual precipitation objects using watershed segmentation. These objects are then linked across consecutive frames using a motion-constrained predictive matching scheme, in which each object's expected position in the subsequent frame is extrapolated from its previous displacement and matched to the best candidate. The resulting time-ordered object sequences are defined as precipitation tracks. Eight representative metrics are characterized to describe precipitation-track kinematics and lifecycle evolution: (1) Lifetime, (2) Track Length, (3) Straightness, (4) Speed, (5) Direction, (6) Turning Rate, (7) ΔArea, and (8) ΔMax Intensity.

Beyond these scalar metrics, two additional track-based analyses are performed. First, the stage-wise lifecycle evolution of tracked storms is analyzed to assess whether their typical growth-to-decay behavior is realistically reproduced. To align storms with different lifetimes onto a common evolutionary axis, each valid track is normalized into five stages from initiation to dissipation, after which stage-wise aggregates are computed to yield composite evolution curves of within-storm mean intensity, area, core area fraction, and propagation speed (Guilloteau and Foufoula-Georgiou, 2024). Second, matched-case trajectory comparisons are presented, providing case-based diagnostics of representative storms, with particular emphasis on moderate-to-high-intensity cases (Kossaifi et al., 2026). Slightly more permissive tracking settings are applied here to favor more continuous tracking of the selected major storms for clearer case-based trajectory visualizations. In each selected window, a major reference track is first selected from MRMS tracks by jointly considering lifetime, spatial extent, and intensity. Tracks are then identified independently in the other datasets, and the one with the largest frame overlap and smallest mean centroid distance to the selected MRMS track is taken as the corresponding match. The centroid trajectories of matched tracks are then compared to assess path fidelity, spread, and displacement.

**Compact Scalar Metrics for Baseline and Ablation Comparisons** A compact set of representative scalar metrics is used to provide concise comparisons across baseline models and ablation settings, spanning complementary aspects of model performance: (1) Continuous Ranked Probability Score (CRPS) for ensemble probabilistic accuracy, (2) Angular Error (AE) and (3) End-Point Error (EPE) for optical-flow-based motion direction and displacement fidelity, (4) Fréchet Video Distance (FVD) for video-level spatiotemporal realism, (5) Wasserstein Distance (WD) for marginal intensity distributional shift, and (6) threshold-weighted Continuous Ranked Probability Score (twCRPS) for upper-tail probabilistic accuracy. Details of these metrics are provided in Appendix D.4.

**Spatial Transferability Test Metrics** For the spatial transferability test, we retain the distributional and temporal-structure diagnostics used in the main evaluation. However, since the high-RQI validation pixels in the NW region are spatially fragmented, spatial organization is characterized using the intensity semivariogram rather than the spectral-/object-based metrics used elsewhere, which require more spatially continuous fields. Details of the semivariogram are provided in Appendix D.5.

## 4. Results

### 4.1. Case Study

We begin with a representative case study to visually assess the storm-scale spatiotemporal dynamics targeted in this study. Figure 3 presents a propagating supercell thunderstorm over the SE region during 20:30 UTC 2 June–06:00 UTC 3 June 2024, encompassing organized initiation, upscale growth, and domain-scale translation within a single sampled clip. Five representative frames are shown for MRMS, IMERG, the three baselines (Attention UNet, DDPM, and ResDDPM), and HiDFiP (a1–f5). The accompanying Hovmöller diagrams further provide compact longitude–time and latitude–time summaries, enabling efficient diagnosis of propagation continuity and displacement (a6–f7).

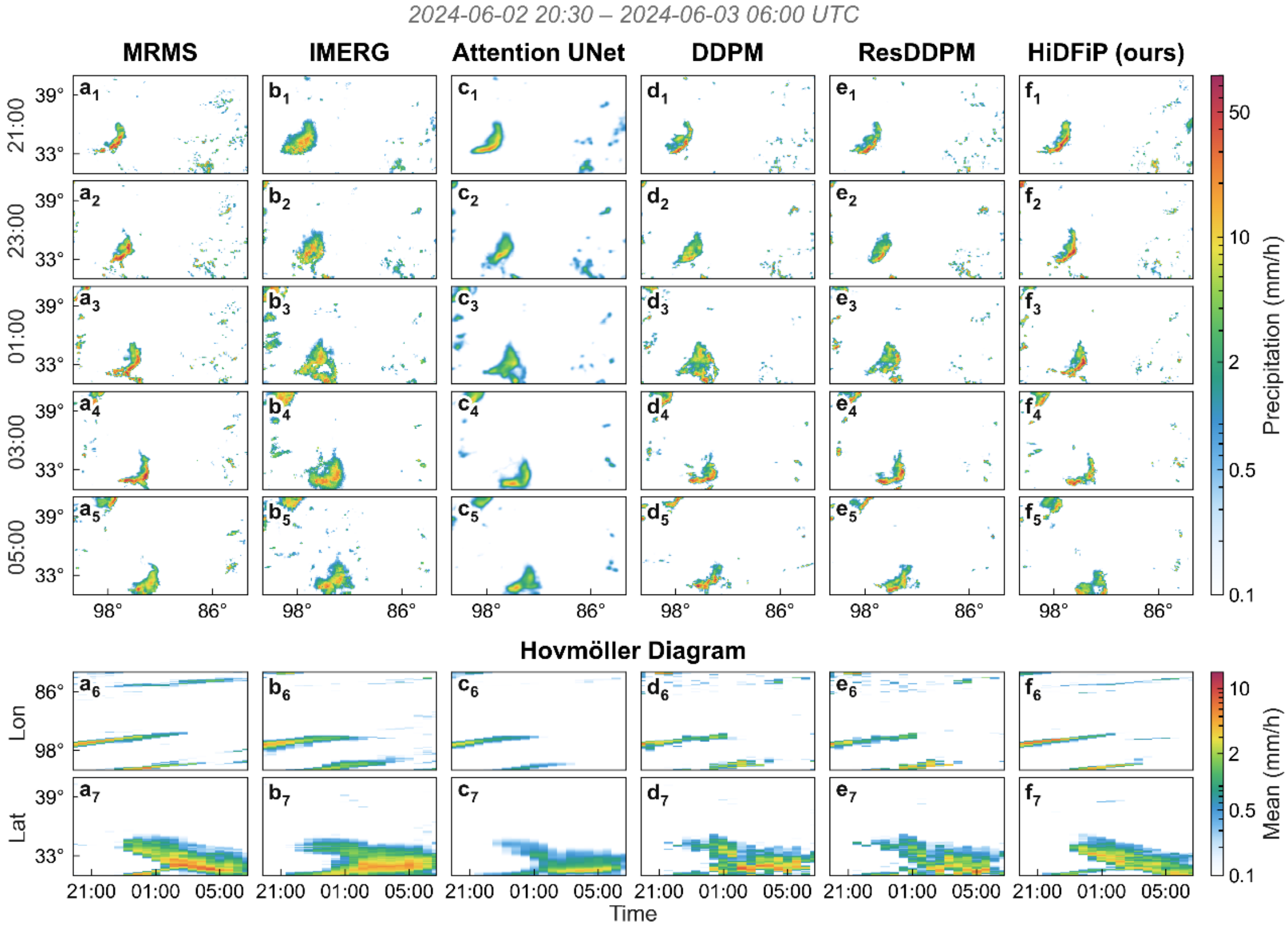

**Figure 3** Case study comparison during 20:30 UTC 2 June–06:00 UTC 3 June 2024 over the Southeast study region (31°–41°N, 102°–82°W), when a propagating supercell thunderstorm develops and traverses the domain. (a1–f5) Five representative half-hourly precipitation snapshots are shown for visualization from MRMS, IMERG, three baseline models (Attention UNet, DDPM, and ResDDPM), and HiDFiP. For the probabilistic models (DDPM, ResDDPM, and HiDFiP), the displayed fields correspond to the first ensemble member. (a6–f7) The corresponding longitude–time and latitude–time Hovmöller diagrams summarize zonal and meridional storm propagation.

Relative to MRMS (a1–a7), IMERG captures the broader precipitating episode but retains the expected dynamical deficiencies of satellite retrievals including inflated and geometrically distorted structures and a relatively smooth yet misplaced evolution trajectory (b1–b7). This discrepancy is especially evident in the latitude–time Hovmöller panel ($b_7$), where the intense precipitation core shows little realistic meridional displacement. The deterministic Attention UNet partially corrects the excessive spatial extent but does so at the expense of diffusing compact high-intensity cores and weakening mesoscale organization, with negligible improvement in temporal dynamics (c1–c7). The image-based DDPM (d1–d7) and ResDDPM (e1–e7) recover finer stochastic texture, but their temporal coherence further degrades and even falls below that of IMERG and Attention UNet, as evidenced by noisier and less organized Hovmöller streaks (d6–d7, e6–e7), partly consistent with the frame-to-frame discontinuity induced by generative stochasticity as schematized in Figure 1.

By contrast, HiDFiP more faithfully reproduces the evolving morphology and translation of the supercell (f1–f7). Across snapshots (f1–f5), its fields remain structurally connected in a manner much closer to MRMS, preserving both the large-scale storm envelope and embedded high-intensity features without the flickering characteristic of frame-independent generation, although residual differences in the exact placement and intensity of some precipitation maxima remain, partly due to the stochastic sampling inherent to generative models. This advantage is equally clear in the Hovmöller diagnostics (f6–f7), where HiDFiP yields smoother, more continuous propagation signals that more closely align with the observed trajectory.

### 4.2. Distributional and Probabilistic Evaluation

Following the representative case study, Figure 4 examines the distributional realism and ensemble calibration of HiDFiP to assess the probabilistic accuracy learned by the model at the pixel level. In the intensity PDF (Figure 4a), HiDFiP closely follows MRMS across most intensities and markedly improves upon IMERG in the heavy-rain tail. Even the ensemble-mean (HiDFiP-EM), despite collapsing the probabilistic ensemble to a deterministic estimate, remains much closer to MRMS than IMERG, suggesting a systematic refinement in the representation of heavy precipitation. The rank histogram is nearly uniform, with slight edge enhancement (Figure 4b), suggesting generally appropriate ensemble spread with only mild underdispersion. The attribute diagram (Figure 4c, top) lies close to the 1:1 line overall, indicating good occurrence calibration. Only a few high-probability bins fall slightly below the diagonal, implying mild overconfidence for near-certain forecasts. The near-zero reliability component (0.0005) and positive resolution term (0.0306) further show that the ensemble has minimal calibration error and retains useful discrimination beyond climatology, respectively. Consistently, the sharpness histogram (Figure 4c, bottom) concentrates probability mass near 0 and, to a lesser extent, near 1, indicating an informative rather than artificially diffuse ensemble. Overall, HiDFiP reproduces a realistic precipitation distribution while providing well-calibrated and informative probabilistic estimates.

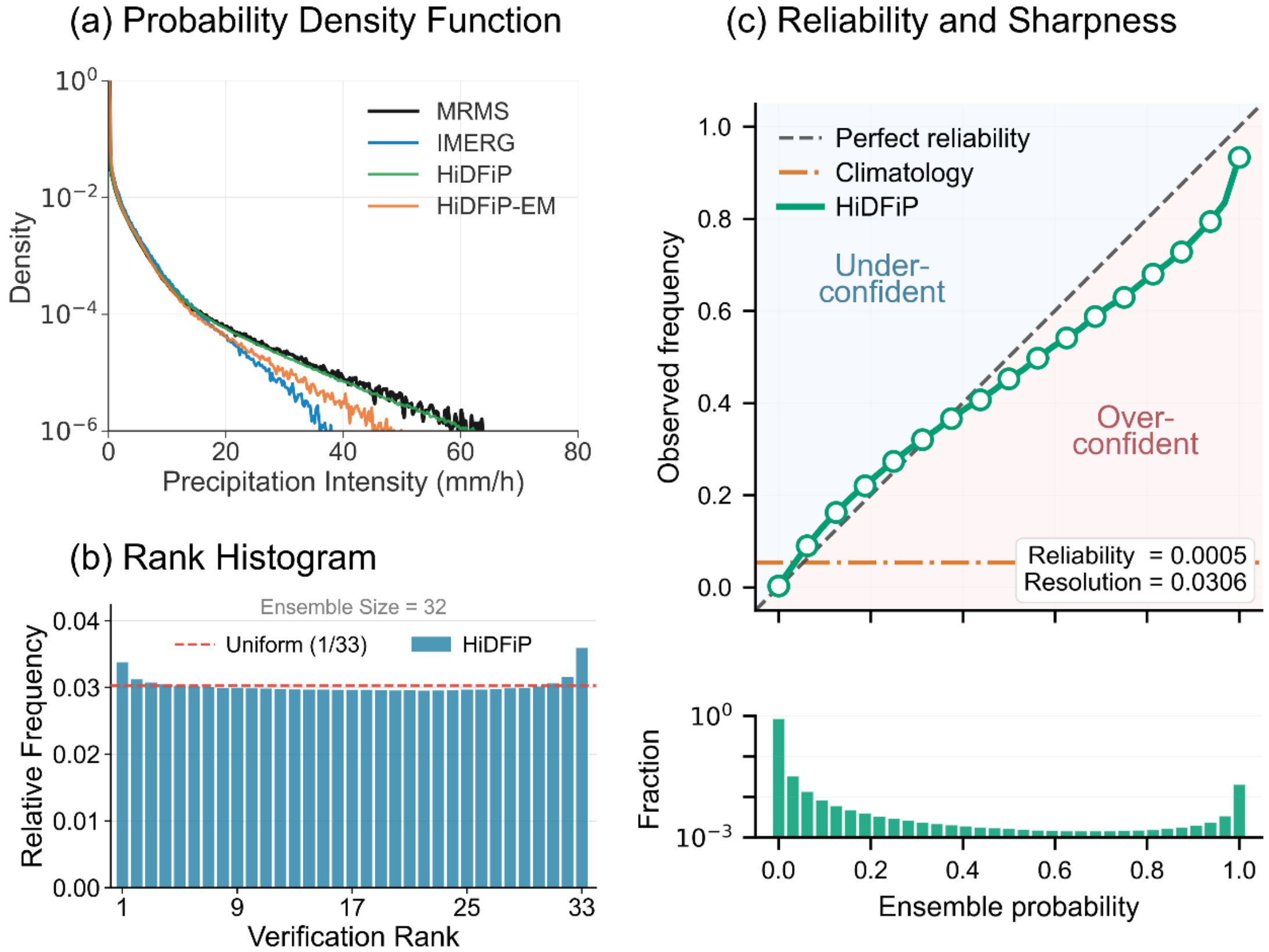

**Figure 4** Distributional and probabilistic verification of HiDFiP precipitation ensembles. (a) Empirical PDFs of half-hourly precipitation intensity for MRMS, IMERG, HiDFiP, and the HiDFiP Ensemble-Mean (HiDFiP-EM), computed with 0.25 mm/h intensity bins and shown on a logarithmic scale to emphasize tail behavior. (b) Rank histogram for the 32-member HiDFiP ensemble (33 verification ranks); the dashed line denotes the uniform frequency for a perfectly reliable ensemble. (c) (top) Attribute diagram for predicted precipitation occurrence (≥0.1 mm/h), where the 1:1 line indicates perfect reliability, and the horizontal line denotes the climatological frequency approximated by the fraction of rainy samples in the test set. Shaded regions mark under- (blue) and over-confidence (red). The inset reports Reliability and Resolution scores from the Brier Score decomposition. (bottom) Sharpness histogram shows the distribution of ensemble probability. Both panels use the same discrete ensemble-probability bins.

### 4.3. Spectral Diagnostics

We next quantitatively examine the spatiotemporal dynamical behavior of the products, starting with the nonparametric multiscale spectral analysis. Figure 5 first presents one-dimensional marginalized spectral diagnostics of precipitation fields in the temporal, zonal, and meridional dimensions, through PSD, spectral coherence, and phase difference relative to MRMS. For clarity, only ResDDPM is shown as the representative baseline, and the same convention is adopted in the subsequent figures.

In the marginalized PSDs (Figure 5a–c), IMERG exhibits a pronounced loss of power at shorter periods and smaller zonal and meridional wavelengths, indicating overly smoothed precipitation variability relative to MRMS. ResDDPM recovers the spatial spectra reasonably well, reflecting its good image-level performance, but retains a clear bias in the temporal spectrum, again highlighting the limitation of image-based models in representing temporal dynamics. By contrast, HiDFiP closely matches MRMS and faithfully reproduces the observed variance distribution across all three dimensions. For spectral coherence (Figure 5d–f), deterministic models may hold a slight advantage because they are not subject to sampling stochasticity. Accordingly, ResDDPM is somewhat weaker than IMERG, whereas HiDFiP, despite being probabilistic, still exceeds IMERG, further underscoring its stronger multiscale consistency with MRMS. More notably, both HiDFiP and ResDDPM substantially reduce the intrinsic spatiotemporal phase biases in IMERG (Figure 5g–i), with HiDFiP still performing slightly better. This suggests that diffusion-based reconstruction, aided by environmental conditioning, improves not only local intensities and overall spatiotemporal organization, but also part of the systematic position/timing offsets inherent in satellite retrievals, such as parallax shift and cloud-top-to-surface representativeness mismatch (Guilloteau et al., 2018).

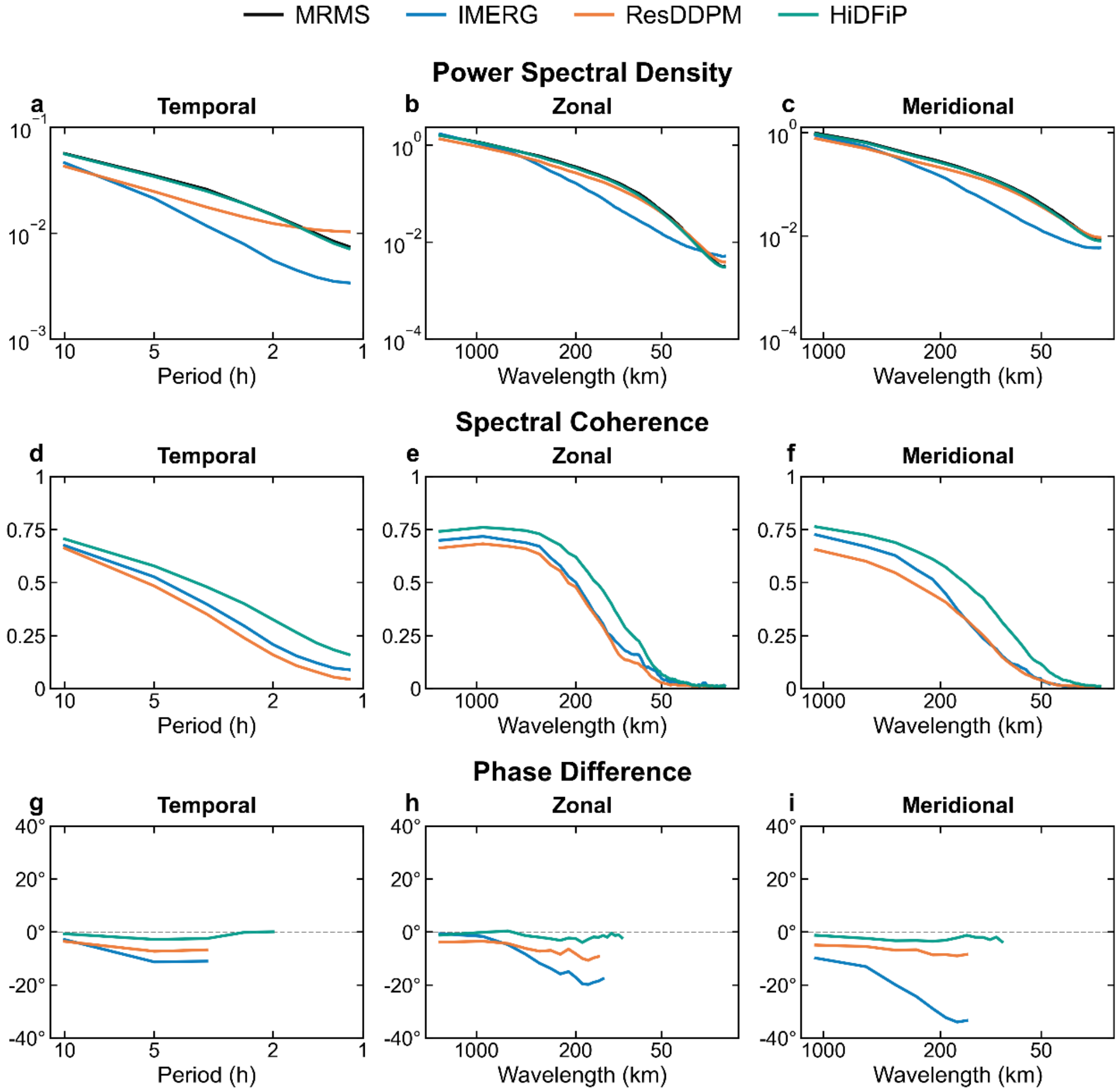

**Figure 5** One-dimensional marginalized spectral diagnostics of IMERG (blue), ResDDPM (orange), and HiDFiP (green) compared to MRMS (black) in temporal, zonal, and meridional dimensions, as a function of temporal period and zonal/meridional spatial wavelength: (a–c) PSD; (d–f) Spectral coherence between each dataset and MRMS; (g–i) Phase difference relative to MRMS, with the dashed line indicating zero phase lag. Phase differences are shown only for coherence > 0.3 to ensure interpretability.

Figure 6 further characterizes spectral behavior of each dataset in joint space–time and space–space representations, facilitating a multiscale assessment of storm propagation and morphology. In MRMS, the wavenumber–frequency ($k$–$\omega$) spectra exhibit clear directional asymmetry, with dominant power concentrated along the eastward and northward branches and along relatively narrow phase-speed rays (Figure 6a–b), indicating a prevailing, approximately non-dispersive northeastward propagation of storms. The wavenumber–wavenumber ($k_x$–$k_y$) spectrum likewise exhibits a pronounced anisotropic lobe that suggests a preferred northeast-tilted storm morphology (Figure 6c). IMERG reproduces the eastward propagation signal and the northeast-oriented spatial structure, although with markedly weakened power at smaller scales (Figure 6d,f). More critically, it largely fails to capture the northward propagation branch (Figure 6e). As expected, ResDDPM recovers the spatial anisotropy more successfully (Figure 6i), yet its $k$–$\omega$ spectra are much more diffuse and show little distinct propagation structure (Figure 6g,h). By contrast, HiDFiP most closely matches MRMS across all three joint spectra, recovering both the eastward and northward propagation preferences and the northeast-tilted spatial organization, with only limited residual smoothing (Figure 6j–l).

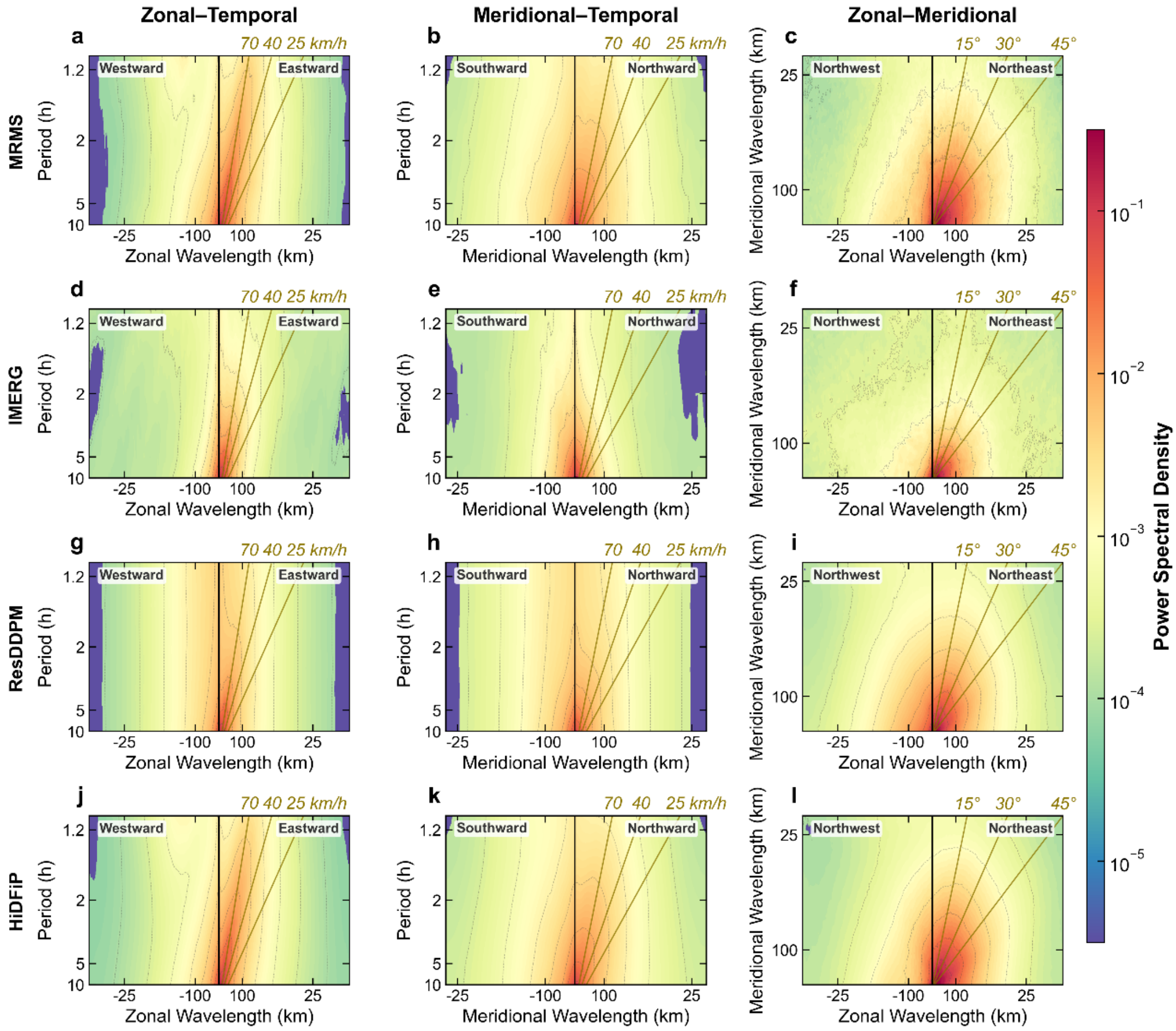

**Figure 6** Wavenumber–frequency ($k$–$\omega$) and wavenumber–wavenumber ($k_x$–$k_y$) spectra illustrating multiscale storm propagation and morphology, revealed by two-dimensional joint PSDs, for (a–c) MRMS, (d–f) IMERG, (g–i) ResDDPM, and (j–l) HiDFiP. For readability, axes are reported as signed wavelength (km) and period (h). In $k-\omega$ space, negative (positive) wavelengths indicate westward/southward (eastward/northward) propagation, and rays mark constant phase speeds, given by wavelength divided by period. In $k_x$–$k_y$ space, negative (positive) zonal wavelengths indicate northwest- (northeast-) tilted structures, and rays mark constant orientation angles, determined by the ratio of meridional to zonal wavelength. Colors indicate logarithmic PSD, and thin contours denote PSD isolines.

For a more intuitive summary of spectrally inferred dynamical behavior, Figure 7 condenses the full space–time spectra into net propagation vectors across four mesoscale wavelength bands. In MRMS, storm propagation is distinctly northeastward at the meso-β1&β2 scales (Figure 7a,b), then gradually rotates toward a more zonal eastward direction at meso-α1&α2 scales (Figure 7c,d), with propagation speed increasing with wavelength. IMERG shows reasonable agreement only at the largest scale (Figure 7d), but exhibits clear biases in both direction and magnitude at the other scales (Figure 7a–c), with its overly zonal vectors consistent with the spectral results in

Figure 6d–f. By contrast, ResDDPM captures the propagation direction reasonably well from meso-β2 to meso-α2, yet consistently underestimates the vector magnitude (Figure 7b–d). At the smallest meso-β1 scale (Figure 7a), ResDDPM shows little coherent direction, indicating its unresolved small-scale dynamics. HiDFiP, in contrast, remains closest to MRMS across all four bands, nearly overlaps with MRMS at meso-β1&β2 (Figure 7a,b), and also closely reproduces the larger-scale eastward propagation at meso-α1&α2 (Figure 7c,d), underscoring its superior fidelity in representing multiscale storm dynamics.

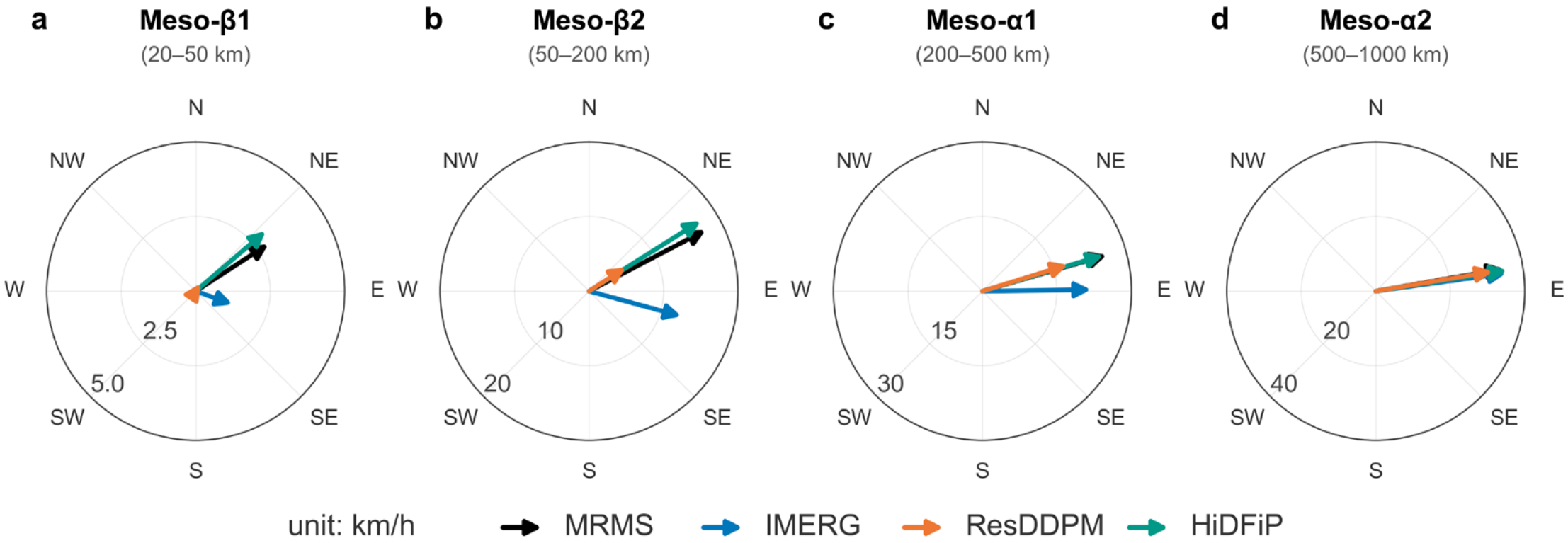

**Figure 7** PSD-weighted net propagation vectors of storms inferred from the space–time spectra for MRMS (black), raw IMERG (blue), ResDDPM (orange), and HiDFiP (green), summarized across four mesoscale wavelength bands: (a) Meso-β1, 20–50 km; (b) Meso-β2, 50–200 km; (c) Meso-α1, 200–500 km; and (d) Meso-α2, 500–1000 km. Arrow azimuth indicates the dominant direction of propagation, and arrow length denotes the PSD-weighted net propagation speed (km/h).

Taken together, the spectral results consistently show that HiDFiP most faithfully reproduces the multiscale precipitation dynamics observed in MRMS, including variance, coherence, phase alignment, directional propagation, and anisotropic morphology, highlighting the advantage of video-based generation in capturing spectrally consistent storm dynamics.

### 4.4. Process-oriented Diagnostics

Complementing the nonparametric spectral analysis, Figure 8 shifts the evaluation toward an object-based, process-oriented framework, treating precipitation as discrete entities and interrogating their attributes across three physically meaningful dimensions, namely precipitation events, systems, and tracks, to provide a more application-relevant view of dataset dynamical behavior. At the event level (Figure 8a), the contrast is immediately apparent: both IMERG and ResDDPM depart markedly from MRMS across multiple attributes, whereas HiDFiP is nearly superposed on the reference. In practical terms, IMERG turns rainfall into too many yet weakened episodes, inflating event counts (47.1/grid/month in IMERG versus 31.5/grid/month in MRMS) while suppressing depth, duration, intensity, and especially short extremes (7.6% in IMERG versus 25.4% in MRMS), consistent with its temporally fragmented and over-smoothed rainfall features. ResDDPM, despite its improved recovery of rainfall intensity and extremes, remains broadly similar to IMERG in overall event behavior and even shows stronger temporal fragmentation, resulting in more frequent events with shorter durations (59.4/grid/month and 1.03 h in ResDDPM, versus 31.5/grid/month and 1.56 h in MRMS). By contrast, HiDFiP preserves the observed balance among event frequency, persistence, intensity, and intermittency, underscoring its high fidelity at the half-hourly event scale.

The system diagnostics reveal a different contrast (Figure 8b). IMERG still produces storms that are systematically too diffuse, with smaller scale, weaker aspect ratio, lower concentration and radial decay, and, most notably, a severely depleted core-area fraction (1.2% in IMERG versus 4.0% in MRMS), indicating a marked failure to recover the internal structure of storm systems. ResDDPM performs markedly better here than at the event level, with the remaining deficiencies largely confined to an underrepresented core area (2.6% versus 4.0% in MRMS) and modest residual biases in aspect ratio and asymmetry, indicating that its storm systems remain somewhat too rounded and internally too homogenized. This may still be related to the frame-independent formulation relaxing the spatiotemporal constraints essential for dynamically plausible morphology. By contrast, HiDFiP remains closely aligned with MRMS across nearly all attributes, except for moderate area overestimation. Given the otherwise strong agreement in morphological metrics, this bias most likely reflects a slight outward extension of weak rain that does not materially alter the overall structure.

The sharpest discrimination emerges at the track level, arguably the most demanding test of coupled spatiotemporal coherence (Figure 8c). IMERG reproduces broad temporal persistence and large-scale path structure (i.e., straightness, overall direction), but still favors tracks that are too long, too fast, and insufficiently curved at local scales (i.e., underestimated turning rate). Its overly large $\Delta$Area but strongly muted $\Delta I_{\max}$ further indicate unstable storm-footprint evolution and overly damped variability in the embedded intense cores. ResDDPM degrades even further, which is physically unsurprising: excessive frame-to-frame jitter makes small-scale objects extremely difficult to track continuously, leaving the derived tracks dominated by larger yet still erratic objects. As a result, track length, speed, turning rate, $\Delta$Area, and $\Delta I_{\max}$ are all substantially exaggerated, whereas lifetime is biased low because tracks break too readily. This behavior shows, in the most direct way, the inability of temporally uncoupled generative models to sustain realistic storm evolution. By contrast, HiDFiP still reproduces the observed track dynamics well across most metrics. Its only notable departure is elevated variability in area and peak intensity, suggesting that its performance still retains a limited imprint of generative stochasticity or residual imperfections from subsequent refinement and tuning.

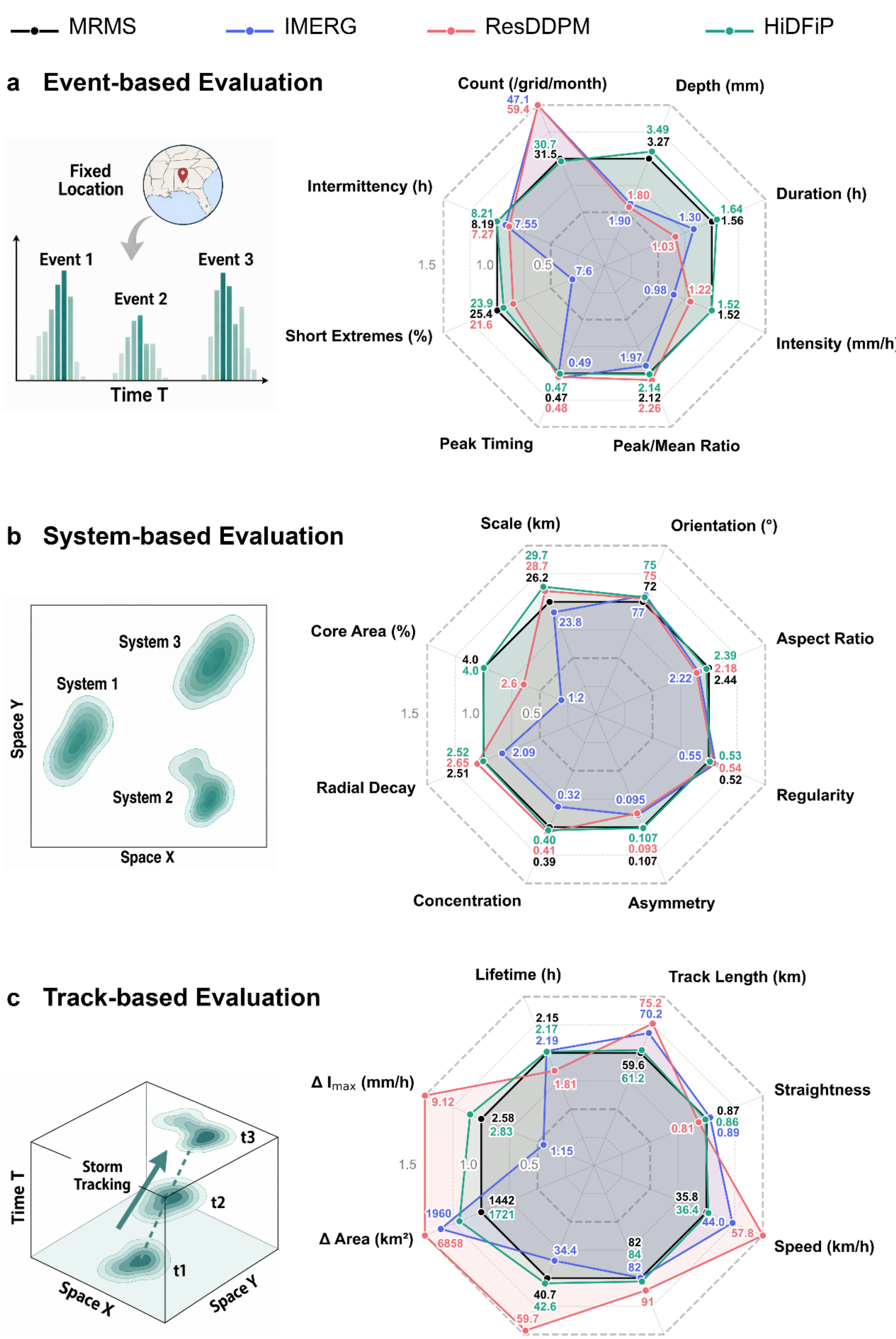

**Figure 8** Process-oriented diagnostics of precipitation objects at three complementary levels: (a) events (1D temporally continuous rainy series at fixed location), (b) systems

(2D spatially continuous rainy area at fixed time), and (c) tracks (3D spatiotemporally continuous rainy objects evolving across space and time). Left panels schematize each object definition, and right radar plots compare the corresponding attributes from MRMS, IMERG, ResDDPM, and HiDFiP. Radar radii are normalized by the MRMS reference for cross-metric comparison, while colored numbers denote the corresponding physical values. Gray numbers indicate the multiplicative factor of each polygon ring relative to the MRMS reference. For visual clarity, values exceeding 1.5 are plotted on the outer 1.5 ring.

Figure 9 extends the evaluation of precipitation tracks beyond scalar metrics by examining the lifecycle evolution of several storm attributes whose temporal variations carry clear dynamical meaning: within-storm mean intensity, area, core fraction, and propagation speed. In MRMS, all these attributes exhibit distinct lifecycle signatures consistent with physical expectations: mean intensity peaks in the early-to-middle lifecycle stages, following the classical storm development pattern (stage B; Figure 9a), propagation speed increases and then plateaus as storms transition from localized growth to a mature propagating stage (Figure 9d), whereas area and core fraction follow a more symmetric growth–decay pattern (Figures 9b,c). IMERG reproduces area evolution reasonably well (Figure 9b), but misses the early-to-midlife intensification peak (Figure 9a), fails to capture core development (Figure 9c), and does not recover the late-life speed plateau (Figure 9d), consistent with the core limitations of satellite precipitation products. ResDDPM, likely because of the previously noted track fragmentation and frame-to-frame jitter, substantially overestimates all four attributes throughout the lifecycle. By contrast, HiDFiP largely reproduces the observed evolution across all metrics, reinforcing its superior dynamical fidelity.

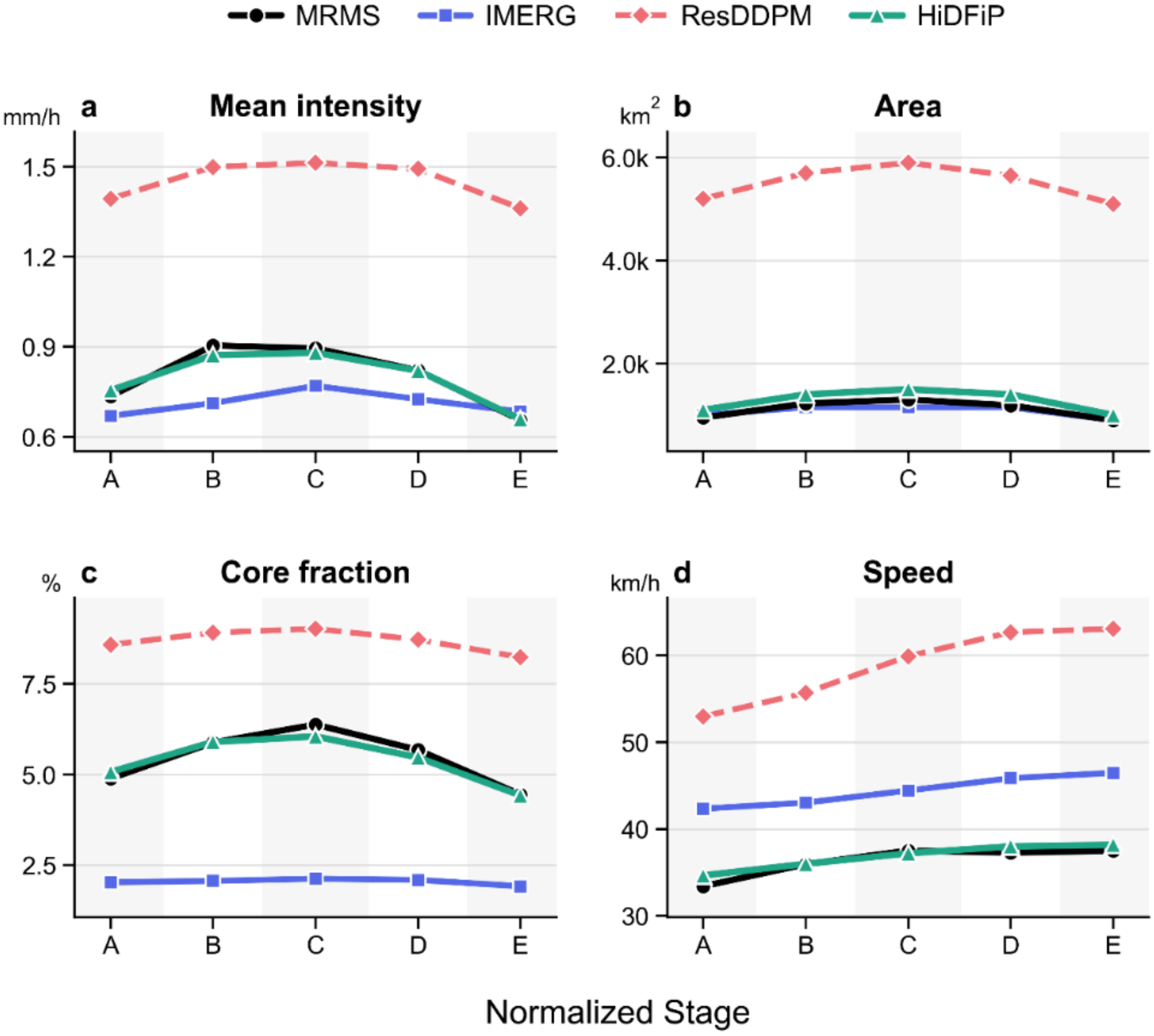

**Figure 9** Five-stage normalized lifecycle evolution of precipitation object characteristics along precipitation tracks for MRMS, IMERG, ResDDPM, and HiDFiP: (a) within-storm mean intensity, (b) area, (c) core area fraction, and (d) propagation speed. Median aggregation is used for (a) and (b) because their stronger cross-storm skewness makes them more sensitive to a few exceptionally intense or large storms, whereas mean aggregation is used for (c) and (d), for which skewness is much weaker.

Figure 10 provides a complementary, case-based view of storm motion by showing centroid trajectories for three representative moderate-to-heavy storm cases spanning three distinct precipitation regimes: synoptic (a winter storm, Figure 10a), convective (a supercell thunderstorm, Figure 10d), and tropical cyclone (Hurricane "Helene", Figure 10g). IMERG exhibits the largest trajectory errors. Although its centroid path agrees reasonably well with MRMS in Case 2 (Figure 10e,f), it shows substantial spatial displacement in Cases 1 and 3 (Figure 10b,c,h,i). This more demanding metric, which requires not only marginal fidelity of individual fields but also concurrent storm evolution in space and time, more clearly exposes IMERG's structural inaccuracy, even though the apparent trajectory discrepancies here may arise from the combined effects of spatiotemporal offsets, within-storm intensity-distribution errors, and track-topology errors, rather than an exaggerated bulk displacement of the storm as a whole. The results therefore raise a clear caution against using satellite precipitation products for track-based analyses of organized precipitating systems, e.g., tropical cyclones and mesoscale convective systems. ResDDPM places its trajectories broadly within the correct spatial envelope, but the individual paths remain visibly noisier, with stochastic initiation and termination, truncated lifecycles, greater local path roughness, and broader member-to-member spread (Figure 10b,e,h). By contrast, despite residual uncertainty and displacement, HiDFiP preserves a coherent and dynamically consistent evolution: when IMERG already follows the MRMS trajectory closely (Case 2), HiDFiP maintains this favorable behavior with a more compact ensemble spread around the IMERG–MRMS paths; when IMERG deviates markedly from MRMS (Cases 1 and 3), HiDFiP substantially steers the ensemble trajectories into closer alignment with the MRMS reference path (Figure 10c,f,i). Taken together, HiDFiP remains the closest overall match to MRMS, extending this agreement beyond aggregate statistics to the concurrent spatiotemporal evolution of individual storms.

Overall, the process-oriented diagnostics show that HiDFiP closely reproduces the precipitation behavior observed in MRMS across events, systems, and tracks, while also faithfully capturing storm evolution profiles and matched storm trajectories. Together, these results suggest that HiDFiP preserves the coupled behavior of storms in space and time, yielding a physically coherent reconstruction of storm evolution.

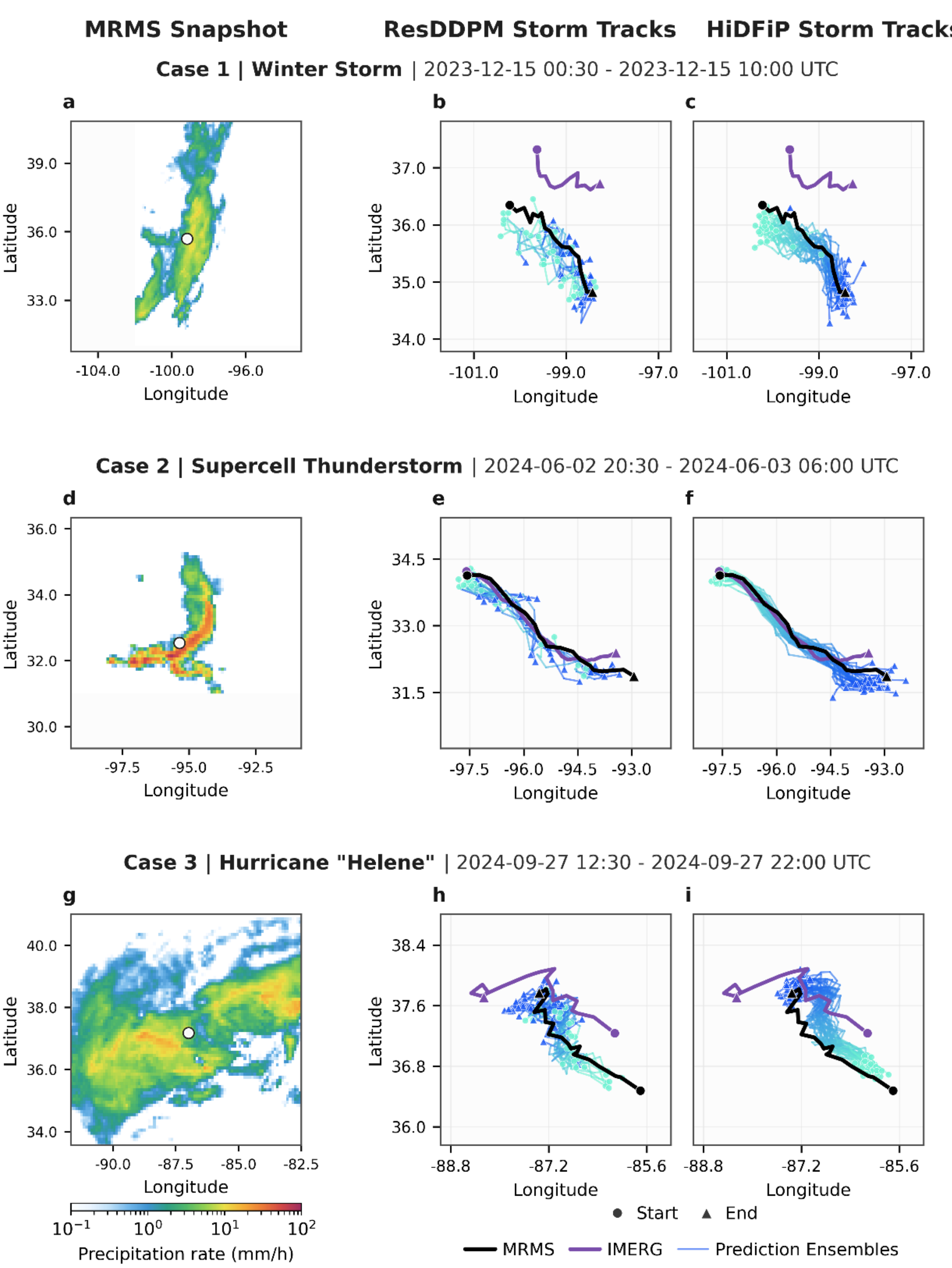

**Figure 10** Centroid trajectories of precipitation tracks within sampled windows for three representative storm events: (a–c) a winter storm (2023-12-15 00:30–10:00 UTC), (d–f) a supercell thunderstorm (2024-06-02 20:30–2024-06-03 06:00 UTC), and (g–i) Hurricane “Helene” (2024-09-27 12:30–22:00 UTC). For each window, corresponding tracks across datasets were identified by anchoring on the MRMS reference track and matching IMERG, HiDFiP, and ResDDPM counterparts based on their spatiotemporal agreement. In each row, the left panel shows the MRMS snapshot at the central frame (frame 11) of the sampled window, recentered on the target precipitation object with its centroid fixed at the image center (white dot), providing visual context for the storm

structure. The middle and right panels show storm trajectories from all ResDDPM and HiDFiP ensemble members (blue/cyan), respectively, together with MRMS (black) and IMERG (purple) tracks. Circles and triangles denote the start and end locations of each track, respectively, and ensemble-member trajectories are additionally colored from cyan to blue from start to end for clarity.

### 4.5. Baseline Comparison and Ablation Study

In this section, we provide a more granular comparison of HiDFiP against IMERG and all the baseline models, while also reporting physically grouped ablations of the ~40 environmental inputs and IMERG itself to isolate the contribution of different variable families. To ensure comparability across experiments, we focus on a compact set of highly informative scalar metrics that summarize complementary aspects of performance. In the baseline comparison (Table S1), HiDFiP performs best overall across the six selected metrics, achieving the lowest CRPS, AE, EPE, FVD, and WD, and a twCRPS essentially tied with the best baseline. This pattern again emphasizes HiDFiP's substantial improvement across complementary and partly competing criteria, including ensemble accuracy, motion fidelity, spatiotemporal realism, distributional agreement, and extremes.

From the ablation test table (Table 1), satellite precipitation remains the indispensable anchor: removing it produces by far the largest degradation. On the other hand, the degraded model still does not collapse entirely (e.g., CRPS remains lower than that of IMERG, and WD stays close to the IMERG level, Table S1), indicating that the ERA5 environmental fields alone still provide substantial predictive constraint (note that ERA5 precipitation itself is not included among the inputs). The converse experiment makes this even clearer: removing all environmental variables leads to broad degradation across all metrics, directly confirming their overall contribution. The category-wise ablations, however, reveal a more nuanced pattern. For example, once sufficiently rich dynamic predictors are included, static topography contributes less than expected, although this is also partly due to the relatively modest relief of the SE region. By contrast, dynamic (e.g., zonal and meridional winds, vertical velocity), thermodynamic (e.g., 2 m temperature, CAPE), and moisture-related variables (e.g., total column water vapor, relative humidity) clearly improve pixel-based metrics such as probabilistic accuracy, distributional fidelity, and extremes, but exert slightly mixed effects on structural spatiotemporal metrics. This likely reflects their role in compensating for the limited direct observational precision of satellite precipitation, while their flow vectors or moisture plumes do not always coincide exactly with the realized evolution of rain systems. Cloud properties (e.g., cloud liquid water content, cloud ice water content), by contrast, more clearly improve dynamical structure, suggesting a more direct link to precipitation organization and occurrence, albeit with slight penalties in ensemble distribution and extremes. Overall, the 4D environmental fields provide an unambiguous net benefit once exploited by the spatiotemporal model, although the heterogeneous ablation responses also suggest that further gains will require more selective and physically informed use of these inputs.

**Table 1** Input ablation study of HiDFiP across six representative scalar evaluation metrics.

| Input Configuration | **CRPS**↓ ($10^{-2}$) (ensemble accuracy) | **AE**↓ (motion direction) | **EPE**↓ ($10^{-2}$) (motion displacement) | **FVD**↓ ($10^{1}$) (spatiotemporal realism) | **WD**↓ ($10^{-2}$) (distributional shift) | **twCRPS**↓ ($10^{-2}$) (tail accuracy) |
|---|---|---|---|---|---|---|
| **All variables** | 7.58 | 1.12 | 3.58 | 1.02 | 0.26 | 3.66 |
| w/o satellite precipitation | 9.59 (+2.01) | 2.14 (+1.02) | 6.91 (+3.33) | 1.62 (+0.60) | 3.09 (+2.83) | 4.42 (+0.76) |
| w/o all environmental variables | 8.08 (+0.50) | 1.25 (+0.13) | 4.00 (+0.42) | 1.15 (+0.13) | 0.59 (+0.33) | 3.79 (+0.13) |
| w/o topography | 7.47 (−0.11) | 1.12 (0.00) | 3.61 (+0.03) | 1.07 (+0.05) | 0.24 (−0.02) | 3.59 (−0.07) |
| w/o dynamics | 7.74 (+0.16) | 1.11 (−0.01) | 3.54 (−0.04) | 1.03 (+0.01) | 0.75 (+0.49) | 3.69 (+0.03) |
| w/o thermodynamics | 8.18 (+0.60) | 1.07 (−0.05) | 3.41 (−0.17) | 0.96 (−0.06) | 1.42 (+1.16) | 3.92 (+0.26) |
| w/o moisture | 8.10 (+0.52) | 1.06 (−0.06) | 3.40 (−0.18) | 1.00 (−0.02) | 1.21 (+0.95) | 3.90 (+0.24) |
| w/o cloud properties | 7.42 (−0.16) | 1.18 (+0.06) | 3.82 (+0.24) | 1.08 (+0.06) | 0.38 (+0.12) | 3.58 (−0.08) |

**Note:** The arrow and parenthetical factor next to each metric indicate, respectively, the direction of better performance and the scaling applied to the reported values. Parenthetical values following each score denote signed differences relative to the "All variables" configuration. "w/o" denotes removal of the corresponding input group.

### 4.6. Spatial Transferability Results

Figure 11 provides a stringent test of spatial transferability by applying the SE-trained model directly to the unseen NW region and evaluating it across the selected metrics, including PDF, temporal PSD, event properties, and semivariogram, with the semivariogram serving as a surrogate metric for spatial structure under the fragmented valid-pixel coverage. All evaluations are restricted to high-MRMS RQI (>0.8) pixels to ensure a reliable reference. Overall, HiDFiP retains meaningful skill under direct spatial transfer and still delivers broad improvements over IMERG. This is evidenced by a more faithful reproduction of multiscale temporal variability, with a temporal PSD closer to MRMS than IMERG (Figure 11b); more realistic event characteristics, with the event-property polygon shifting toward MRMS and showing the same broad direction of correction relative to IMERG in event count, duration, intensity, and depth as in the SE region (Figure 11c); and a more realistic spatial autocorrelation structure, with a semivariogram closer to MRMS (Figure 11d). The main residual bias manifests as a wet bias in moderate-to-heavy precipitation (~10–30 mm/h) (Figure 11a), excessive spectral power (Figure 11b), and an inflated fraction of short-extreme events (Figure 11c). A simple QM post-processing step further improves several aspects of performance, bringing event properties into near alignment (Figure 11c), and improving the semivariogram overall (Figure 11d). Its effect on the PDF and PSD, however, is more mixed: QM improves moderate rainfall (~10–20 mm/h) and longer timescales (>~5 h), but suppresses heavier rates (>~20 mm/h), damps short-timescale variability (<~2 h), and slightly underrepresents short-duration extremes, although it still remains superior to IMERG across these aspects. This experiment partially emulates the gains that could be achieved in completely radar-free regions through a simple gauge-based distributional correction.

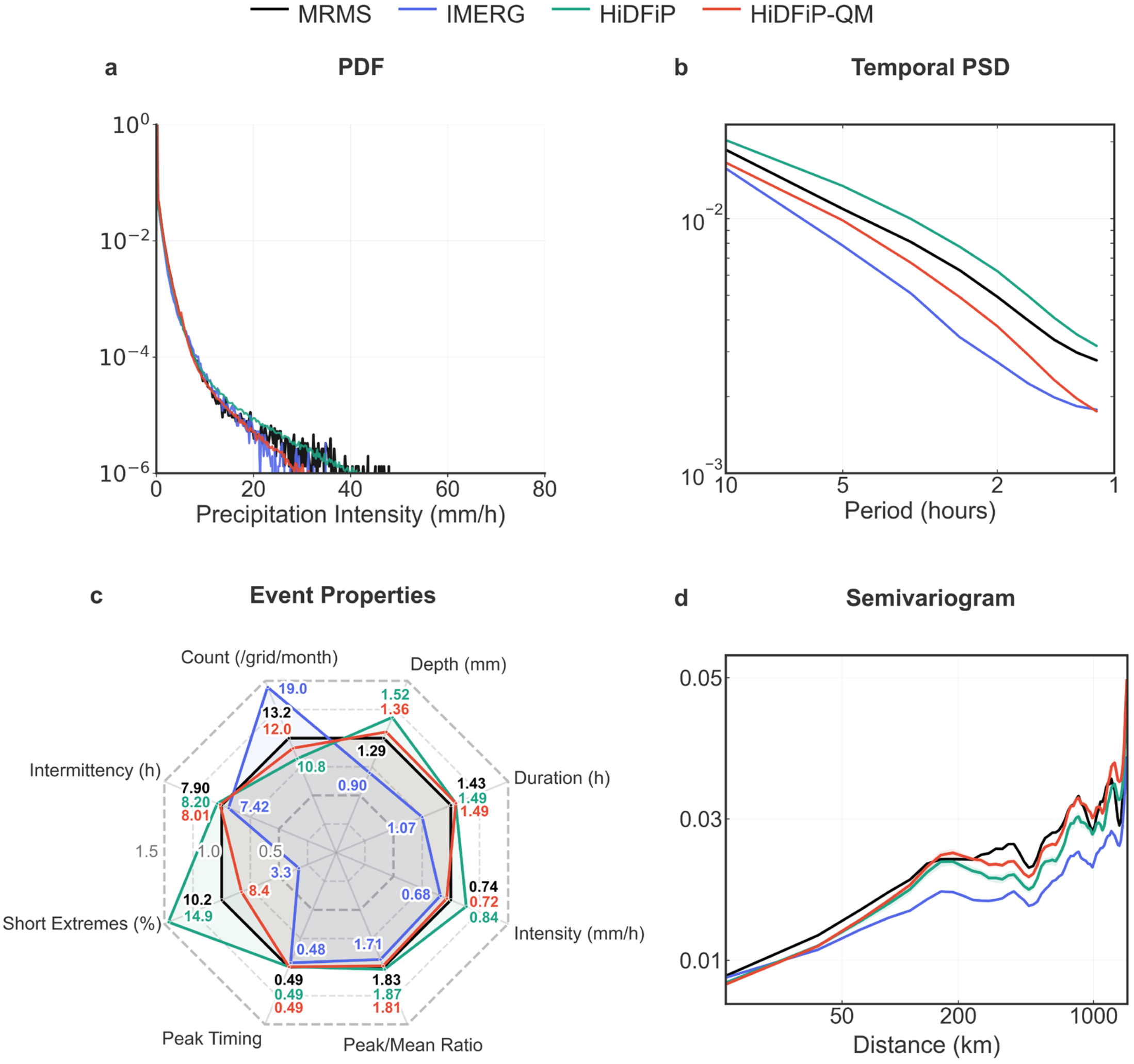

**Figure 11** Results of the spatial transferability experiment over the NW study region during the October 2023–September 2024 test period, using the model trained over the SE region. Selected metrics include (a) PDF, (b) temporal PSD, (c) precipitation event properties, and (d) semivariogram for MRMS (black), IMERG (blue), HiDFiP (green), and HiDFiP-QM (red), with HiDFiP-QM denoting HiDFiP after 5-fold cross-validated

quantile mapping against MRMS. All statistics and the quantile mapping are computed only over pixels with high MRMS RQI (>0.8) to ensure a reliable reference.

In summary, by combining explicit spatiotemporal modeling with physically informed four-dimensional environmental constraints, HiDFiP retains meaningful skill under a deliberately demanding, adaptation-free transfer from the SE to the climatically contrasting NW, rather than behaving as a region-specific statistical remapping. Relative to MRMS, residual discrepancies remain moderate, whereas gains over IMERG are retained across most diagnostics. This native transferability points to substantial value for reconstructing high-fidelity precipitation fields in low-RQI, effectively radar-void regions. It also highlights the potential of extending the framework with dedicated transfer-learning strategies to produce more reliable precipitation products in radar-sparse regions whose climatic regimes could differ markedly from those of the training domain.

## 5. Discussion

A key implication of this study is that explicit temporal modeling is not optional for high-resolution satellite precipitation products. This is made clear by the image-based diffusion baselines: although they reproduce many frame-wise spatial metrics reasonably well, their temporal fidelity deteriorates substantially. Unlike AI forecasting emulation/bias-correction tasks, which are naturally grounded in temporal autoregressive generation or physical dynamical models, or land surface temperature/aerosol optical depth inversion tasks, which rely on continuous geostationary observations, the temporal information available for precipitation retrievals is unavoidably constrained by the day-scale revisit intervals of the polar-orbiting observations on which such retrievals rely. Such weak observational constraints, together with precipitation's intrinsically high variability and the stochasticity of generative sampling, make satellite-based precipitation estimation especially susceptible to frame-to-frame instability when each time step is treated independently. Temporal consistency is therefore the central organizing principle of high-resolution satellite precipitation product development, not merely an add-on to frame-wise estimation, thereby making video-based models likely indispensable for such tasks in the future.

An equally important implication concerns evaluation itself. As this study shows, comprehensive evaluation of multifaceted spatiotemporal dynamics is indispensable: strong performance on conventional yet partial metrics can still mask important deficiencies under more stringent and diverse tests, especially in process-oriented diagnostics central to flood forecasting, convective-system diagnosis, and climate-model evaluation. Developed in tandem with the video-diffusion model, the hierarchical evaluation suite presented here therefore establishes a validation paradigm for AI-based satellite precipitation products, with HiDFiP's near-MRMS agreement across these complementary diagnostics providing strong evidence of genuine spatiotemporal realism. Taken together, this emphasis on spatiotemporal dynamical fidelity—both in modeling and in evaluation—constitutes the central conceptual contribution of this study and provides a unifying framework for the development and assessment of next-generation satellite precipitation products.

The encouraging direct-transfer performance across contrasting climate regimes suggests that, through joint spatiotemporal modeling and the extraction of 4D physical information from a large set of environmental predictors, HiDFiP can learn generalizable precipitation–environment relationships, providing a plausible route toward future global product development. The modest remaining gap relative to in-domain evaluation, however, indicates that dedicated transfer-learning/domain-adaptation strategies, together with broader evaluation across diverse climate regimes, will still be required to reach a production-grade global product. A related implication concerns the use of environmental variables. The heterogeneous ablation responses across predictor categories suggest that these inputs are better viewed not as a flat predictor stack, but as structured physical constraints that help regularize an otherwise largely underdetermined inverse problem. This in turn points to future gains from more selective, physically informed conditioning strategies, especially those that explicitly encode multiscale spatiotemporal dependencies and lagged relationships among predictors (Mu et al., 2025).

Several limitations remain, but they also point directly to important future developments. First, this study has not yet implemented fully continuous long-video generation (e.g., year-scale sequences), which is itself a nontrivial topic. However, the stable spatiotemporal consistency achieved within 20-frame video units suggests that the core short- to medium-range dynamical reconstruction problem has been substantially addressed. Given that the input predictors themselves retain reasonable temporal coherence over longer timescales, extending generation to longer sequences should be tractable (Aich et al., 2026; Qian et al., 2026). Future work can extend HiDFiP toward longer continuous precipitation sequences through techniques such as hierarchical temporal completion, overlap-based segment stitching, and context-aware sequential extension (Henschel et al., 2025; Qiu et al., 2024; Yin et al., 2023). In addition, although HiDFiP performs well across the vast majority of diagnostics, a few residual limitations remain, including slight ensemble underdispersion, mildly overestimated precipitation-system area, and modest spatial offsets in matched storm tracks. These remaining gaps suggest room for further gains from both more physically informed, domain-knowledge-integrated model designs, e.g., advection-/continuity-aware evolution modules, conservation-constrained learning, and structured physical conditioning (Metzl et al., 2025; Wang et al., 2024a; Zhang et al., 2023b), and advances in the broader video-diffusion literature, e.g., joint space–time denoising architectures, explicit motion conditioning, and motion-aware training objectives (Bar-Tal et al., 2024; Guo et al., 2024; Wang et al., 2024b).

Finally, it should also be noted that this study builds its satellite-based precipitation estimation from the Level 3 seamlessly gridded product rather than from native brightness temperatures. This design choice is reasonable, because an input already closer to the target field provides a more informative conditional prior and makes learning more tractable, especially under limited data, even though it inherits biases introduced during product generation. Importantly, the sparse and irregular sampling of native brightness-temperature swaths does not preclude video-based modeling; rather, these observations can be treated as partially observed conditioning signals in mask-aware or inpainting-style video diffusion frameworks (Lugmayr et al., 2022; Voleti et al., 2022). Future work should nevertheless explore direct conditioning on native brightness temperatures, which could preserve information lost during downstream retrieval and thus offer a higher performance ceiling, albeit through a substantially more challenging inverse problem.

## 6. Conclusions

In this study, we develop a video-diffusion framework for satellite-based High-Dynamical-Fidelity Precipitation (HiDFiP) field generation. The framework is motivated by two challenges: (1) reconstructing precipitation with high dynamical fidelity beyond the limited spatiotemporal coverage of ground-radar, and (2) addressing a persistent weakness of high-resolution satellite precipitation products—their intrinsic temporal unreliability arising from the inhomogeneity, intermittency, and indirectness of passive-microwave retrievals. IMERG serves as the primary satellite precipitation input, complemented by ~40 four-dimensional atmospheric/land-surface variables from ERA5/ERA5-Land to compensate for the information deficit in satellite retrievals, while MRMS provides the ground-radar QPE reference for training and CONUS serves as the testbed for both temporal extrapolation and spatial transfer experiments. A comprehensive diagnostic suite is carefully developed to interrogate the multifaceted dynamical fidelity of the generated fields, spanning case studies, distributional/ensemble diagnostics, spectral analyses, process-oriented evaluations at the precipitation-event, -system, and -track levels, and compact scalar metrics.

Across these complementary evaluations, HiDFiP consistently remains the closest match to MRMS, outperforming both IMERG, which exhibits broad dynamical-fidelity deficiencies, and the image-wise diffusion baselines, which show particularly weak temporal consistency. Specifically, it performs strongly in precipitation distributional fidelity, ensemble calibration, upper-tail skill, spectral power and coherence, storm timing/location, directional propagation, event episodicity and intermittency, precipitation-system morphology and internal structure, and storm-track kinematics and lifecycle evolution. Ablation studies further confirm that these environmental predictors provide substantial skill gains that vary across physically distinct predictor categories, rather than merely playing a supplementary role. Moreover, HiDFiP still performs reasonably well when the model is applied directly to an unseen region, suggesting that the framework learns transferable process-level relationships rather than merely a local statistical remapping. Taken together, this study advances a new video-diffusion paradigm for satellite-based, high-dynamical-fidelity precipitation field generation and positions temporal dynamical consistency as a central requirement for next-generation satellite-based precipitation products. By coupling spatiotemporal generation with process-oriented evaluation, the proposed framework provides both an algorithmic and diagnostic foundation for the future development of long-term, global, radar-grade precipitation records.

## Acknowledgements

This work was supported by NASA (Grants 80NSSC23K1304 and 80NSSC26K1141), the National Science Foundation (Grant IIS2324008), and the Samueli endowed chair to Efi Foufoula-Georgiou.

## Appendices

### Appendix A. Environmental Variables Used in HiDFiP

- **Topography:** surface geopotential from ERA5-Land.
- **Dynamics:** surface pressure and 10 m zonal/meridional winds from ERA5-Land; mean sea-level pressure, geopotential height at 850/500/200 hPa, zonal and meridional winds at 850/500/200 hPa, vertical velocity at 850/500/200 hPa, and horizontal divergence at 850/500/200 hPa from ERA5.
- **Thermodynamics:** 2 m temperature and skin temperature from ERA5-Land; air temperature at 850/500/200 hPa, convective available potential energy, and K index from ERA5.
- **Moisture:** 2 m dewpoint temperature and volumetric soil water in layer 1 from ERA5-Land; specific humidity at 850/500/200 hPa, total column water vapor, and vertically integrated moisture divergence from ERA5.
- **Cloud properties:** total cloud cover, total column cloud liquid water, and total column cloud ice water from ERA5.

### Appendix B. Detailed Internal Model Architecture

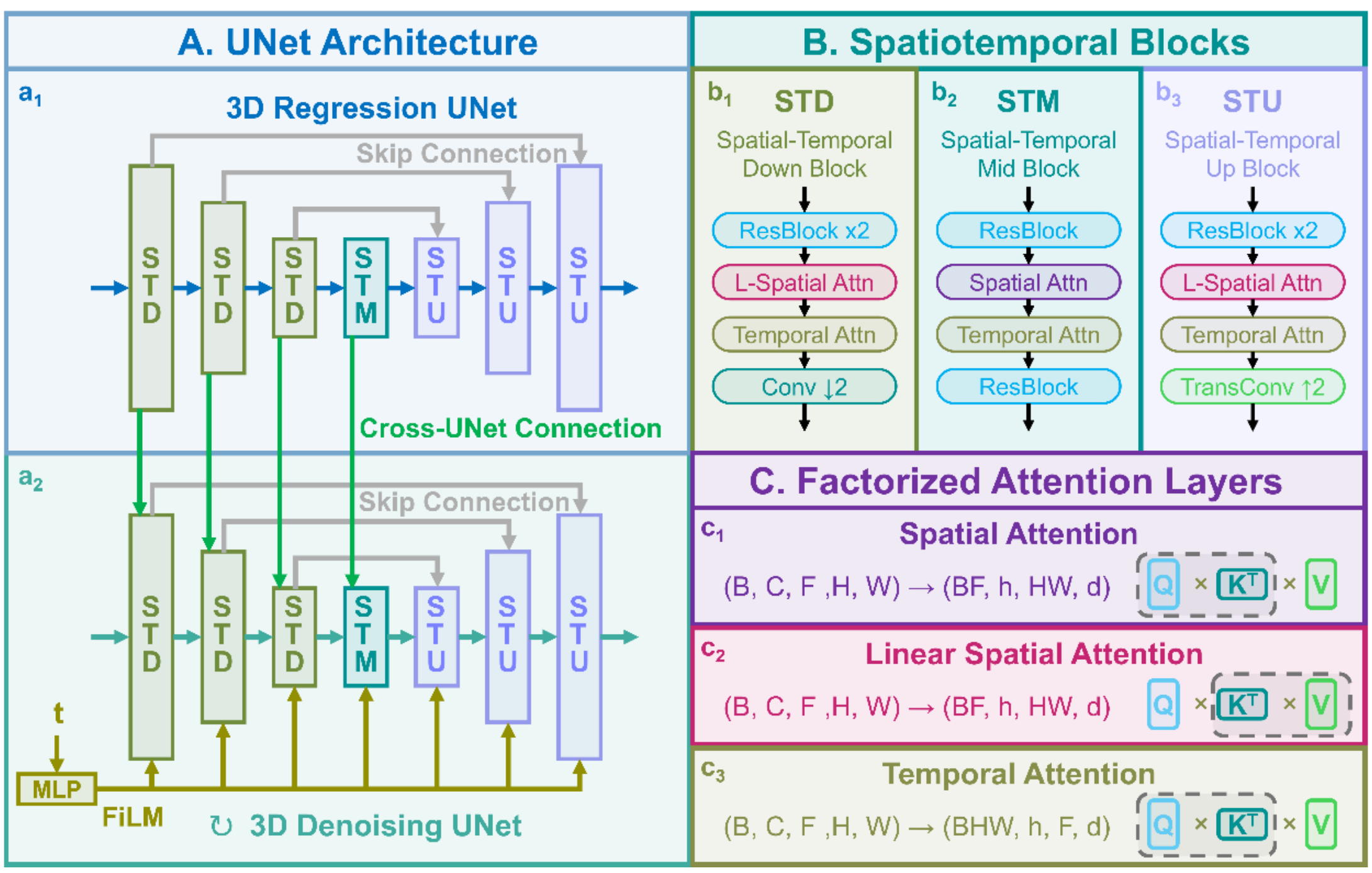

**Figure B.1.** Detailed internal architecture of the model. **A. UNet Architecture**: Paired 3D UNets form the spatiotemporal backbone and connect mean prediction with diffusion refinement. **B. Spatiotemporal Blocks**: Modular down/mid/up blocks implement the multiscale feature transformations used in the 3D UNets. **C. Factorized Attention Layers**: Attention is factorized into separate spatial and temporal operations on intermediate 5D features.

**UNet Architecture** Panel A illustrates the architectures of the two UNets. Both the R-UNet and D-UNet adopt a standard 3D encoder-decoder with skip connections, in which each resolution level comprises spatiotemporal down/mid/up blocks (STD/STM/STU) tailored to video-formatted tensors. To strengthen conditioning during diffusion, multiscale encoder features from the R-UNet are injected into the D-UNet at matched resolutions, providing deterministic context consistent with the mean estimate. The diffusion timestep $t$ is embedded by a Multi-Layer Perceptron (MLP) and injected into the D-UNet via Feature-wise Linear Modulation (FiLM), following standard practice (Perez et al., 2018).

**Spatiotemporal Blocks** Panel B details the STD/STM/STU blocks. Each block combines residual blocks (ResBlocks) with factorized spatial–temporal attention (Attn), a widely used design in video models that avoids the prohibitive cost of full 3D attention (Ho et al., 2022a; Ho et al., 2022b). STD/STU use (transposed) convolutions for resolution transitions and linear spatial attention to reduce computation, whereas the bottleneck STM retains standard spatial attention to maximize representational capacity at the coarsest scale.

**Factorized Attention Layers** Panel C illustrates the operation of the factorized attention layers. For an intermediate feature map $\boldsymbol{z} \in \mathbb{R}^{B\times C\times F\times H\times W}$, spatial attention is computed frame-wise by folding temporal frames into the batch: $\boldsymbol{z}^{\mathrm{sp}} \in \mathbb{R}^{BF\times h\times HW\times d}$. Accordingly, temporal attention is computed grid-wise by folding spatial locations into the batch: $\boldsymbol{z}^{\mathrm{te}} \in \mathbb{R}^{BHW\times h\times F\times d}$, where $h$ is the number of attention heads and $d$ is the per-head dimension. Additionally, linear attention adopted in STD/STU employs kernelization to reorder the Query–Key–Value (Q–K–V) computations, providing an efficient approximation to full attention at substantially lower computational cost (Katharopoulos et al., 2020).

## Appendix C. Full Video Diffusion Formulation and Loss Function

### C.1. Video Diffusion Formulation

**Structured Noise Prior** Following Ge et al. (2023), let $\boldsymbol{\epsilon} = \left[\boldsymbol{\epsilon}^1; \dots; \boldsymbol{\epsilon}^i; \dots; \boldsymbol{\epsilon}^F\right] \in \mathbb{R}^{FHW}$ denote the noise maps $\boldsymbol{\epsilon}^i \in \mathbb{R}^{HW}$ stacked over $F$ frames. For each frame $i$, $\boldsymbol{\epsilon}^i$ is constructed as a variance-preserving mixture of a frame-shared component, $\boldsymbol{\epsilon}_{\mathrm{shared}} \sim \mathcal{N}\left(\mathbf{0}, \frac{\lambda^2}{1+\lambda^2}\mathbf{I}_{HW}\right)$, and a frame-wise independent component, $\boldsymbol{\epsilon}_{\mathrm{ind}}^i \sim \mathcal{N}\left(\mathbf{0}, \frac{1}{1+\lambda^2}\mathbf{I}_{HW}\right)$:

$$\boldsymbol{\epsilon}^i = \boldsymbol{\epsilon}_{\mathrm{shared}} + \boldsymbol{\epsilon}_{\mathrm{ind}}^i, \tag{C.1}$$

where $\mathbf{I}_{HW}$ is the identity matrix over the per-frame grids (i.e., spatial dimension remains i.i.d.). $\{\boldsymbol{\epsilon}_{\mathrm{ind}}^i\}_{i=1}^F$ are independent across frames and independent of $\boldsymbol{\epsilon}_{\mathrm{shared}}$. $\lambda$ controls their relative contribution (set to 1 for balanced mixing). This construction preserves the per-frame marginal covariance while inducing a constant cross-frame covariance:

$$\mathrm{Cov}(\boldsymbol{\epsilon}^i, \boldsymbol{\epsilon}^j) = \begin{cases} \mathbf{I}_{HW}, & i = j \\ \frac{\lambda^2}{1+\lambda^2}\mathbf{I}_{HW}, & i \neq j \end{cases}, \tag{C.2}$$

thereby uniquely specifying a joint Gaussian distribution from which the noise injected throughout the diffusion process is sampled:

$$\boldsymbol{\epsilon} \sim \mathcal{N}(\mathbf{0}, \boldsymbol{\Sigma}),\ \boldsymbol{\Sigma} = \left[\mathrm{Cov}(\boldsymbol{\epsilon}^i, \boldsymbol{\epsilon}^j)\right]_{i,j=1}^{F} \in \mathbb{R}^{FHW \times FHW}. \tag{C.3}$$

**Diffusion Processes** Following the DDPM forward noising formulation, we replace the identity covariance $\mathbf{I}$ with the structured noise covariance $\boldsymbol{\Sigma}$ defined above. For each residual target $\boldsymbol{r}_0$, the forward process gradually corrupts the residual by adding noise:

$$q(\boldsymbol{r}_t \mid \boldsymbol{r}_{t-1}) = \mathcal{N}(\boldsymbol{r}_t; \sqrt{\alpha_t}\,\boldsymbol{r}_{t-1}, \beta_t \boldsymbol{\Sigma}),\ \alpha_t = 1 - \beta_t, \tag{C.4}$$

where $\{\beta_t\}_{t=1}^{T}$ is a predefined variance schedule and $T$ denotes the total number of DDPM diffusion steps. Let $\bar{\alpha}_t = \prod_{s=1}^{t} \alpha_s$, with $\bar{\alpha}_0 = 1$ by convention. The forward-process marginal distribution then admits the closed form:

$$q(\boldsymbol{r}_t \mid \boldsymbol{r}_0) = \mathcal{N}(\boldsymbol{r}_t; \sqrt{\bar{\alpha}_t}\boldsymbol{r}_0, (1 - \bar{\alpha}_t)\boldsymbol{\Sigma}), \tag{C.5}$$

which enables direct sampling via reparameterization,

$$\boldsymbol{r}_t = \sqrt{\bar{\alpha}_t}\boldsymbol{r}_0 + \sqrt{1 - \bar{\alpha}_t}\boldsymbol{\epsilon},\ \boldsymbol{\epsilon} \sim \mathcal{N}(\mathbf{0}, \boldsymbol{\Sigma}). \tag{C.6}$$

The DDIM-based reverse process reuses the same structured-covariance forward marginals in (C.5) while traversing the reduced sampling schedule $\{t_k\}_{k=0}^{K}$ in reverse, from $t_K = T$ to $t_0 = 0$. For each update $t_k \to t_{k-1}$, the corresponding marginal-preserving DDIM denoising transition is

$$q_\sigma(\boldsymbol{r}_{t_{k-1}} \mid \boldsymbol{r}_{t_k}, \boldsymbol{r}_0) = \mathcal{N}\left(\left(\sqrt{\bar{\alpha}_{t_{k-1}}} - \sqrt{\bar{\alpha}_{t_k}}\sqrt{\frac{1-\bar{\alpha}_{t_{k-1}}-\sigma_{t_k,t_{k-1}}^2}{1-\bar{\alpha}_{t_k}}}\right)\boldsymbol{r}_0 + \sqrt{\frac{1-\bar{\alpha}_{t_{k-1}}-\sigma_{t_k,t_{k-1}}^2}{1-\bar{\alpha}_{t_k}}}\boldsymbol{r}_{t_k}, \sigma_{t_k,t_{k-1}}^2\boldsymbol{\Sigma}\right), \tag{C.7}$$

where $\sigma_{t_k,t_{k-1}} = \eta\sqrt{\frac{1-\bar{\alpha}_{t_{k-1}}}{1-\bar{\alpha}_{t_k}}}\sqrt{1 - \frac{\bar{\alpha}_{t_k}}{\bar{\alpha}_{t_{k-1}}}}$, and $\eta$ controls the sampling stochasticity. During reverse generation, $\boldsymbol{r}_0$ is intractable, precluding direct use of (C.7) for sampling. However, we can use a denoising neural network (D-UNet) to predict $\boldsymbol{r}_0$, which in turn induces an estimated posterior distribution. Concretely, in the reverse process we learn conditional denoising transitions given $(\boldsymbol{x}, \bar{\boldsymbol{y}}_\theta)$ and denote the estimated clean residual as $\hat{\boldsymbol{r}}_{0|t_k}$. Substituting $\hat{\boldsymbol{r}}_{0|t_k}$ into (C.7) yields the fully parameterized reverse transition:

$$p_\phi(\boldsymbol{r}_{t_{k-1}} \mid \boldsymbol{r}_{t_k}, \boldsymbol{x}, \bar{\boldsymbol{y}}_\theta) = \mathcal{N}\left(\sqrt{\frac{1-\bar{\alpha}_{t_{k-1}}-\sigma_{t_k,t_{k-1}}^2}{1-\bar{\alpha}_{t_k}}}\boldsymbol{r}_{t_k} + \left(\sqrt{\bar{\alpha}_{t_{k-1}}} - \sqrt{\bar{\alpha}_{t_k}}\sqrt{\frac{1-\bar{\alpha}_{t_{k-1}}-\sigma_{t_k,t_{k-1}}^2}{1-\bar{\alpha}_{t_k}}}\right)\hat{\boldsymbol{r}}_{0|t_k}, \sigma_{t_k,t_{k-1}}^2\boldsymbol{\Sigma}\right),$$

$$p(\boldsymbol{r}_T) = \mathcal{N}(\mathbf{0}, \boldsymbol{\Sigma}). \tag{C.8}$$

Iterating (C.8) for $k = K, \ldots, 1$ generates the final residual sample $\hat{\boldsymbol{r}}_0$.

### C.2. Loss Function

**Diffusion Objective** The denoising loss is defined under the $v$-parametrization. The velocity target is given by: $\boldsymbol{v}_t = \sqrt{\bar{\alpha}_t}\boldsymbol{\epsilon} - \sqrt{1-\bar{\alpha}_t}\,\boldsymbol{r}_0$, and we train the D-UNet to predict $\hat{\boldsymbol{v}}_t = \boldsymbol{v}_\phi(\boldsymbol{r}_t, t, \boldsymbol{x}, \bar{\boldsymbol{y}}_\theta)$ by minimizing:

$$\mathcal{L}_v(\theta,\phi) = \mathbb{E}_{t,\epsilon,(x,y)}[\|\boldsymbol{v}_t - \hat{\boldsymbol{v}}_t\|_2^2]. \tag{C.9}$$

This objective is equivalent to recovering $\hat{\boldsymbol{r}}_{0|t}$ required for the reverse transition via the deterministic transformation:

$$\hat{\boldsymbol{r}}_{0|t} = \sqrt{\bar{\alpha}_t}\,\boldsymbol{r}_t - \sqrt{1-\bar{\alpha}_t}\,\hat{\boldsymbol{v}}_t. \tag{C.10}$$

**Wasserstein Distance Regularization** The Wasserstein-distance regularizer is efficiently computed using the Sliced Wasserstein-1 Distance (SWD) (Liu et al., 2025b). Given $\hat{\boldsymbol{r}}_{0|t}$ derived from (C.10), we can recover the predicted precipitation sample at diffusion time $t$: $\hat{\boldsymbol{y}}_{0|t} = \bar{\boldsymbol{y}}_\theta + \hat{\boldsymbol{r}}_{0|t}$. Given minibatches $\boldsymbol{y}, \hat{\boldsymbol{y}}_{0|t} \in \mathbb{R}^{B\times1\times F\times H\times W}$, since batch size $B$ is typically small in video diffusion (here $B=1$), we merge $B$ and $F$ to construct the empirical sample set, treating each frame as one sample in $\mathbb{R}^{HW}$: $\boldsymbol{y}^{\mathrm{vec}} \in \mathbb{R}^{BF\times HW}$, $\hat{\boldsymbol{y}}^{\mathrm{vec}} \in \mathbb{R}^{BF\times HW}$. The sliced Wasserstein distance is defined as the expected 1D Wasserstein distance over random projections $\boldsymbol{u} \sim \mathrm{Unif}(\mathbb{S}^{HW-1})$, where $\mathrm{Unif}(\mathbb{S}^{HW-1})$ denotes the uniform distribution on the unit hypersphere $\mathbb{S}^{HW-1} := \{\boldsymbol{u} \in \mathbb{R}^{HW} : \|\boldsymbol{u}\|_2 = 1\}$:

$$W_1^S(\boldsymbol{y}, \hat{\boldsymbol{y}}_{0|t}) = \mathbb{E}_{\boldsymbol{u}\sim\mathrm{Unif}(\mathbb{S}^{HW-1})}[W_1(\boldsymbol{y}^{\mathrm{vec}}\boldsymbol{u}, \hat{\boldsymbol{y}}^{\mathrm{vec}}\boldsymbol{u})], \tag{C.11}$$

where $W_1$ for empirical samples admits the closed form:

$$W_1(\boldsymbol{y}^{\mathrm{vec}}\boldsymbol{u}, \hat{\boldsymbol{y}}^{\mathrm{vec}}\boldsymbol{u}) = \frac{1}{BF}\sum_{bf=1}^{BF}\left|(\boldsymbol{y}^{\mathrm{vec}}\boldsymbol{u})_{(bf)} - (\hat{\boldsymbol{y}}^{\mathrm{vec}}\boldsymbol{u})_{(bf)}\right|, \tag{C.12}$$

where $(\cdot)_{(bf)}$ denotes the ascending sorting. In practice, we approximate the expectation by averaging over $N_{\mathrm{proj}}$=100 randomly sampled directions. Averaging across diffusion times yields the SWD loss

$$\mathcal{L}_{\mathrm{SWD}}(\theta,\phi) = \mathbb{E}_{t,\epsilon,(x,y)}\left[W_1^S(\boldsymbol{y}, \hat{\boldsymbol{y}}_{0|t})\right], \tag{C.13}$$

which is added to the standard denoising objective as

$$\mathcal{L}(\theta,\phi) = (1-\omega)\mathcal{L}_v(\theta,\phi) + \omega\mathcal{L}_{\mathrm{SWD}}(\theta,\phi),\ \omega \in [0,1], \tag{C.14}$$

where $\omega$ is the trade-off weight between the two loss terms.

## Appendix D. Detailed Evaluation Metric Definitions

### D.1. Distributional/Ensemble Metrics

**Probability Density Function** We estimate the empirical probability density function of precipitation intensity from all evaluated grid cells and time frames. For an intensity bin $B_k = [b_k, b_{k+1})$, the density is estimated as

$$\hat{f}(B_k) = \frac{1}{N\Delta b_k}\sum_{n=1}^{N}\mathbf{1}\{p_n \in B_k\}, \tag{D.1}$$

where $p_n$ denotes precipitation intensity at sample $n$, $N$ is the total number of samples, and $\Delta b_k = b_{k+1} - b_k$ is the bin width. It is used to assess whether each product reproduces the marginal distribution of precipitation intensity, particularly the frequency of heavy rain.

**Rank Histogram** The rank histogram diagnoses the ensemble calibration and dispersion. For each verification instance, the observation $y$ is pooled with the $M$-member ensemble and assigned a verification rank $r \in \{1, \dots, M+1\}$ within the combined ordering. For a perfectly reliable ensemble, the observation and ensemble members are statistically indistinguishable draws from the same predictive distribution, and the rank is uniformly distributed $r \sim \text{Unif}\{1, \dots, M+1\}$. A flat histogram therefore signifies a reliable ensemble, whereas a U-shaped histogram reveals underdispersion, with observations disproportionately falling in the outer ranks.

**Attribute Diagram** An attribute diagram is employed to evaluate the calibration performance of the model (Hsu and Murphy, 1986). For an $M$-member ensemble, samples are grouped into $M+1$ categories according to the predicted probability $p_i \in \{i/M\}_{i=0}^{M}$, where $i$ out of $M$ ensemble members exceed the rainy threshold $\tau = 0.1\ \text{mm/h}$ and the observed frequency $\bar{o}_i$ is computed from the samples within each category. The diagram plots these discrete $p_i$ against the corresponding $\bar{o}_i$, where the 1:1 diagonal denotes perfect calibration. Additionally, based on the Brier Score decomposition (Murphy, 1973), the scalar $\text{Reliability}$ and $\text{Resolution}$ scores are quantified as:

$$\text{REL} = \frac{1}{N}\sum_{i=0}^{M} n_i(\bar{o}_i - p_i)^2, \tag{D.2}$$

and

$$\text{RES} = \frac{1}{N}\sum_{i=0}^{M} n_i(\bar{o}_i - \bar{o})^2, \tag{D.3}$$

respectively, where $\bar{o}$ is the climatological frequency calculated over the entire test set. $\text{REL}$ measures the mean-squared calibration error (lower is better), whereas $\text{RES}$ quantifies discrimination relative to climatology (higher is better).

**Sharpness Histogram** The sharpness histogram reports the sample fraction $n_i/N$ in each probability category, where $n_i$ is the sample size in category $i$ and $N$ is the total sample size. Sharper estimates concentrate more mass near 0 or 1 rather than at intermediate probabilities.

### D.2. Spectral Diagnostics

Spectral diagnostics are computed from the three-dimensional Fourier representation of each precipitation video clip. For each video clip $I(t, y, x)$, the clip-wise spatiotemporal mean $\bar{I}$ is first removed to suppress the zero-frequency component, and a separable Tukey window $w(t, y, x)$ is applied to reduce spectral leakage.

$$\tilde{I}(f, k_y, k_x) = \mathcal{F}_{t,y,x}\{w(t, y, x)\,[I(t, y, x) - \bar{I}]\}, \tag{D.4}$$

where $f$ denotes temporal frequency (cycles per day) and $k_x$, $k_y$ denote zonal and meridional wavenumbers (cycles per degree), respectively. For any two fields $X$ and $Y$, the clip-wise auto- and cross-spectra are

$$S_{XX}(f, k_y, k_x) = |\tilde{X}|^2, \quad S_{XY}(f, k_y, k_x) = \tilde{X}\tilde{Y}^*, \tag{D.5}$$

with $(\cdot)^*$ denoting complex conjugation.

For each retained spectral view, the auto- and cross-spectra are marginalized over the non-retained spectral dimensions. One-dimensional views retain $f$, $k_y$, or $k_x$ corresponding to temporal, meridional, and zonal spectra, respectively, while two-dimensional joint views retain $f$–$k_y$, $f$–$k_x$, or $k_y$–$k_x$ spectra. Denoting the resulting marginalized auto- and cross-spectra again by $S_{XX}$, $S_{YY}$, and $S_{XY}$ for notational simplicity, the power spectral density (PSD) of field $X$ for that spectral view is defined as the clip-averaged auto-spectrum:

$$\mathrm{PSD}_X = \langle S_{XX} \rangle, \tag{D.6}$$

where $\langle \cdot \rangle$ denotes averaging across evaluated clips. Coherence and Phase Difference between fields $X$ and $Y$ are computed as

$$C_{XY} = \frac{|\langle S_{XY} \rangle|}{\sqrt{\langle S_{XX} \rangle \langle S_{YY} \rangle}}, \ \Phi_{XY} = \arg(\langle S_{XY} \rangle), \tag{D.7}$$

### D.3. Process-oriented Diagnostics

#### D.3.1. Representative Precipitation Event Properties

- **Count (/grid/month)** Equivalent monthly event count per grid: $N_{\mathrm{E}}$.
- **Depth (mm)** Event-total precipitation accumulation: $D_{\mathrm{E}}$.
- **Duration (h)** Event timespan: $T_{\mathrm{E}}$.
- **Intensity (mm/h)** Event-mean precipitation rate: $I_{\mathrm{E}}$.
- **Peak/Mean Ratio** Intra-event sharpness: $R_{\mathrm{I}} = \frac{I_{\mathrm{Emax}}}{I_{\mathrm{E}}}$, where $I_{\mathrm{Emax}}$ is the event-peak intensity.
- **Peak Timing** Temporal asymmetry, reported as the normalized peak-time position within the event: $R_{\mathrm{T}} = \frac{T_{\mathrm{Emax}} - T_{\mathrm{E0}}}{T_{\mathrm{E}}}$, where $T_{\mathrm{E0}}$ and $T_{\mathrm{Emax}}$ denote the onset and peak time, respectively.
- **Short Extremes** Fractional contribution of short-duration extreme events ($T_{\mathrm{E}}$<1.5 h, $I_{\mathrm{E}}$>5 mm/h) to the total precipitation amount: $F_{\mathrm{SE}} = \frac{\sum D_{\mathrm{SE}}}{\sum D_{\mathrm{E}}}$, where $D_{\mathrm{SE}}$ denotes the depth of the short-duration extreme event and the summations are taken over all events within the study period.
- **Intermittency (h)** Mean dry-spell duration calculated by averaging the dry-spell duration between consecutive events: $T_{\mathrm{dry}}$, characterizing temporal clustering of events.

#### D.3.2. Representative Precipitation System Properties

- **Scale (km)** Effective linear extent of the system: $L_\mathrm{S} = \sqrt{A_\mathrm{S}} = \sqrt{N_\mathrm{pix} a_\mathrm{pix}}$, where $A_\mathrm{S}$ is the system area, $N_\mathrm{pix}$ is the system pixel count, and $a_\mathrm{pix}$ is the per-pixel area approximated by the grid-cell area at the central latitude of the study region.
- **Orientation (°)** Principal elongation direction of the system, defined as the acute major-axis angle of the fitted ellipse relative to north: $\theta_\mathrm{S}$.
- **Aspect Ratio** Elongation strength of the system, defined as the ratio of major-axis length $l_\mathrm{maj}$ to the minor-axis length $l_\mathrm{min}$ of the fitted ellipse: $AR = \frac{l_\mathrm{maj}}{l_\mathrm{min}}$.
- **Regularity** Shape simplicity, quantified by the intersection-over-union between the system region $\mathcal{S}$ and its fitted ellipse $\mathcal{E}$: $\mathrm{IoU} = \frac{|\mathcal{S} \cap \mathcal{E}|}{|\mathcal{S} \cup \mathcal{E}|}$.
- **Asymmetry** Spatial off-centering of the system core, quantified by the scale-normalized displacement between the geometric centroid $\boldsymbol{c}$ and the intensity-weighted centroid $\boldsymbol{c}_\mathrm{I}$: $\delta_\mathrm{S} = \frac{\|\boldsymbol{c} - \boldsymbol{c}_\mathrm{I}\|_2}{L_\mathrm{S}}$
- **Concentration** Degree to which total precipitation volume is contributed by a limited number of high-intensity pixels, characterized by the Gini coefficient $G_\mathrm{S} = \frac{1}{2N_\mathrm{pix}^2 \mu_\mathrm{pix}} \sum_{i=1}^{N_\mathrm{pix}} \sum_{j=1}^{N_\mathrm{pix}} |x_i - x_j|$, $\mu_\mathrm{pix} = \frac{1}{N_\mathrm{pix}} \sum_{i=1}^{N_\mathrm{pix}} x_i$, where $x_i$ is the precipitation intensity at the $i$-th pixel.
- **Radial Decay** Core-to-periphery intensity attenuation of the system, represented by the fitted coefficient $\alpha_\mathrm{S}$ in the exponential model $I_\mathrm{pix}(r) = I_\mathrm{C} \exp\left(-\alpha_\mathrm{S} \frac{r}{L_\mathrm{S}}\right)$, where $I_\mathrm{pix}(r)$ denotes the pixel intensity at radial distance $r$ from the intensity-weighted centroid $\boldsymbol{c}_\mathrm{I}$, and $I_\mathrm{C}$ is the intensity at $\boldsymbol{c}_\mathrm{I}$.
- **Core Area (%)** Fraction of convective-core area $A_\mathrm{core}$ (area with $I_\mathrm{pix}$>5 mm/h) to the total system area: $F_\mathrm{core} = \frac{A_\mathrm{core}}{A_\mathrm{S}}$.

### D.3.3. Representative Precipitation Track Properties

- **Lifetime (h)** Temporal persistence of the track: $T_\mathrm{K}$.
- **Track Length (km)** Total translation distance of the object centroid over its lifetime: $S_\mathrm{K}$.
- **Straightness** Trajectory directness: $ST_\mathrm{K} = \frac{d_\mathrm{K}}{S_\mathrm{K}}$, where $d_\mathrm{K}$ is the distance between the genesis and lysis object centroids.
- **Speed (km/h)** Mean translation velocity of the track centroid over its lifetime: $V_\mathrm{K} = \frac{S_\mathrm{K}}{T_\mathrm{K}}$.

- **Direction (°)** Mean propagation heading of the track centroid relative to north over its lifetime: $\theta_{\mathrm{K}}$.
- **Turning Rate (°)** Mean absolute frame-to-frame change in direction, reflecting trajectory curvature: $\Delta\theta_{\mathrm{K}}$.
- **ΔArea (km²)** Mean absolute frame-to-frame change in object area, indicating areal growth/decay rate: $\Delta A_{\mathrm{K}}$.
- **ΔMax Intensity (mm/h)** Mean absolute frame-to-frame change in the within-object maximum intensity, capturing intensity evolution dynamics: $\Delta I_{\mathrm{Kmax}}$.

**D.4. Compact Scalar Metrics**

**Continuous Ranked Probability Score (CRPS)** CRPS is a strictly proper scoring rule for ensemble predictions that provides an overall pointwise measure of distributional accuracy by comparing the full predictive distribution with the verifying observation (Gneiting and Raftery, 2007). For a pointwise $M$-member ensemble estimate $\{\hat{y}^{(i)}\}_{i=1}^{M}$ and an observation $y$, CRPS is computed from the empirical distribution as

$$\mathrm{CRPS} = \frac{1}{M}\sum_{i=1}^{M}\left|\hat{y}^{(i)} - y\right| - \frac{1}{2M^2}\sum_{i=1}^{M}\sum_{j=1}^{M}\left|\hat{y}^{(i)} - \hat{y}^{(j)}\right|, \tag{D.8}$$

where $M$ is the ensemble size, $\hat{y}^{(i)}$ is the $i$-th ensemble member, and $y$ is the observation. Lower CRPS indicates a forecast distribution that places probability mass closer to the observation. For deterministic predictions, CRPS reduces to the absolute error $|\hat{y} - y|$ and thus coincides with Mean Absolute Error (MAE) after averaging over all verification instances.

**Angular Error (AE)** AE quantifies directional error in storm motion by measuring the angular deviation between predicted and reference optical-flow vectors derived from consecutive precipitation fields (Baker et al., 2011).

$$\mathrm{AE} = \frac{1}{|\mathcal{V}|}\sum_{(s,t)\in\mathcal{V}} \arccos\left(\frac{\hat{\boldsymbol{m}}_{s,t}\cdot\boldsymbol{m}_{s,t}+1}{\sqrt{\left(\|\hat{\boldsymbol{m}}_{s,t}\|_2^2+1\right)\left(\|\boldsymbol{m}_{s,t}\|_2^2+1\right)}}\right), \tag{D.9}$$

where $\mathcal{V}$ is the set of valid pixel–time locations with $|\mathcal{V}|$ denoting its size. $\hat{\boldsymbol{m}}_{s,t}$ and $\boldsymbol{m}_{s,t}$ are the optical-flow vectors estimated from the predicted and observed precipitation sequences at location $s$ and time $t$, respectively. Lower AE indicates more faithful reproduction of motion direction.

**End-Point Error (EPE)** EPE quantifies the magnitude and displacement error in storm motion by measuring the mean Euclidean distance between predicted and reference optical-flow vectors over valid pixel–time locations (Baker et al., 2011).

$$\mathrm{EPE} = \frac{1}{|\mathcal{V}|}\sum_{(s,t)\in\mathcal{V}}\left\|\hat{\boldsymbol{m}}_{s,t} - \boldsymbol{m}_{s,t}\right\|_2. \tag{D.10}$$

Lower EPE indicates more faithful reproduction of motion magnitude and displacement.

**Fréchet Video Distance (FVD)** FVD is a feature-space distance metric for spatiotemporal realism, quantifying discrepancies between predicted and observed videos in a pretrained representation (Unterthiner et al., 2018). Although originally designed for natural video generation, it can serve here as an auxiliary proxy for the overall three-dimensional spatiotemporal structural fidelity of precipitation fields (Gao et al., 2023).

$$\mathrm{FVD} = \|\mu_{\hat{z}} - \mu_z\|_2^2 + \mathrm{Tr}\left(\Sigma_{\hat{z}} + \Sigma_z - 2\left(\Sigma_z^{1/2}\Sigma_{\hat{z}}\Sigma_z^{1/2}\right)^{1/2}\right), \quad \text{(D.11)}$$

where $\hat{z}$ and $z$ denote feature embeddings extracted from the predicted and observed videos, respectively, and $(\mu_{\hat{z}}, \Sigma_{\hat{z}})$ and $(\mu_z, \Sigma_z)$ denote the corresponding means and covariances of their Gaussian-approximated feature distributions**.** Lower FVD indicates that the generated videos are closer to the reference videos in a joint spatiotemporal feature space.

**Wasserstein Distance (WD)** WD denotes the first-order Wasserstein distance between the empirical marginal precipitation-intensity distributions of the predicted and reference fields:

$$\mathrm{WD} = W_1(\hat{P}, P) = \inf_{\pi\in\Pi(\hat{P},P)} \int_{\mathbb{R}\times\mathbb{R}} |a - b| \; d\pi(a, b), \quad \text{(D.12)}$$

where $\hat{P}$ and $P$ denote the predicted and reference marginal precipitation-intensity distributions, $\Pi(\hat{P}, P)$ is the set of all couplings between them, $\pi$ is a transport plan, and $a$ and $b$ are precipitation-intensity values drawn from the two marginals. Lower WD indicates smaller marginal distributional shift.

**Threshold-weighted Continuous Ranked Probability Score (twCRPS)** twCRPS is a threshold-weighted variant of CRPS for upper-tail verification, emphasizing predictive performance above a prescribed high-intensity threshold. Here, we use the threshold-censoring transformation $\psi_\tau(z) = \max(z, \tau)$, where $\tau = q_{0.99}$ denotes the empirical 99th percentile of the reference precipitation distribution over the evaluation set. The empirical ensemble estimator is:

$$\mathrm{twCRPS}_\tau = \frac{1}{M}\sum_{i=1}^{M}\left|\psi_\tau(\hat{y}^{(i)}) - \psi_\tau(y)\right| - \frac{1}{2M^2}\sum_{i=1}^{M}\sum_{j\neq i}\left|\psi_\tau(\hat{y}^{(i)}) - \psi_\tau(\hat{y}^{(j)})\right|, \quad \text{(D.13)}$$

where $\psi_\tau(\cdot)$ maps all values below $\tau$ to $\tau$, thereby concentrating the score on upper-tail behavior; $M$ is the ensemble size, $\hat{y}^{(i)}$ is the $i$-th ensemble member, and $y$ is the observation. For deterministic predictions, twCRPS reduces pointwise to the threshold-censored absolute error $|\psi_\tau(\hat{y}) - \psi_\tau(y)|$, and is therefore equivalent to the MAE of the transformed variable when averaged over the evaluation set. Lower twCRPS indicates better predictive performance for high-intensity precipitation.

### D.5. Intensity Semivariogram for Spatial Transferability Test

The semivariogram summarizes scale-dependent spatial dissimilarity in log-transformed precipitation intensity. For each distance bin centered at $h$, we compute the isotropic semivariance as:

$$\gamma(h) = \frac{1}{2|T||N(h)|}\sum_{t\in T}\sum_{(i,j)\in N(h)}[z_t(i) - z_t(j)]^2, \;\; z_t(i) = \log(1 + P_t(i)), \quad \text{(D.14)}$$

where $N(h)$ denotes the set of unordered sampled pixel pairs whose separation distances fall within that bin, $T$ is the set of evaluated frames, and $P_t(i)$ is the precipitation intensity at pixel $i$ and frame $t$. The log transform reduces the leverage of rare extreme intensities in the squared pairwise differences. Closer agreement with the MRMS semivariogram indicates a more realistic representation of scale-dependent spatial autocorrelation and precipitation-field texture.

# Supplementary Information for

## Video Diffusion for Satellite-based High-Dynamical-Fidelity Precipitation (HiDFiP) Field Generation

**Contents:**

### Supplementary Method S1. Implementation Details

All experiments are conducted on a single NVIDIA L40S GPU (48 GB). Each training sample is a spatiotemporal clip of size $F \times H \times W = 20 \times 100 \times 200$. To increase the number of training samples, we use sliding-window sampling along the temporal dimension with a stride of one frame. For preprocessing, positively skewed variables (precipitation, Convective Available Potential Energy (CAPE), Total Column Cloud Ice Water (TCIW), Total Column Cloud Liquid Water (TCLW), and all moisture-category variables) are log-transformed and min-max scaled, while the remaining variables are linearly min-max normalized. The diffusion model is trained with a fixed diffusion depth of $T = 1500$, a sigmoid $\beta_t$ schedule, and Min-SNR-$\gamma$ loss reweighting ($\gamma = 5$) (Hang et al., 2023). Both the R- and D-UNets adopt a base channel width of 64 with multipliers (1, 2, 4) across three resolution levels, yielding bottleneck feature maps at a relatively high spatial resolution (25×50). This design avoids over-compressing extreme values while leveraging spatial attention to preserve long-range interactions.

The model is trained using the Adam optimizer for 4,000 updates, with a batch size of 1 and gradient accumulation over 10 steps. A single training run requires approximately four days. The learning rate follows a warmup-cosine schedule: 5% linear warmup from $5\times10^{-7}$ to $1\times10^{-4}$, followed by cosine annealing back to $5\times10^{-7}$. We maintain an exponential moving average of model parameters (decay 0.9999) and use the EMA weights for all reported results. During training, we randomly hold out eight video clips from the training pool as a compact validation set and remove all overlapping samples from the remaining training set to avoid leakage. This design preserves as many training samples as possible while keeping validation overhead manageable, yet still provides a non-overlapping set for monitoring convergence and tuning hyperparameters, broadly consistent with validation practices adopted in related studies (Guilloteau et al., 2025; Srivastava et al., 2024). For ensemble evaluation, we generate 32 members for our method as well as all generative baselines and ablation variants. Due to limited computational resources, the substantial memory and time overhead associated with video-diffusion training and sampling, and the need to conduct multiple baseline comparisons and ablation experiments, we evaluate on a stratified (i.e., sampling one 20-frame clip from every 200-frame interval to ensure coverage across all seasons) 10% subset of the held-out test period (Oct 2023–Sep 2024), comprising 88 clips; this still amounts to 1,126,400,000 ensemble-member gridpoint-time samples. During testing, we employ DDIM sampling with $K = 300$ steps and $\eta = 1$ to retain stochasticity (Song et al., 2021), using quadratic spacing to select the inference timesteps over $[1, T]$. This configuration preserves fidelity while enhancing the sampling of extremes.

**Table S1** Quantitative comparison of IMERG, three reference baselines, and HiDFiP (ours) against MRMS across six representative scalar evaluation metrics.

| **Dataset/Model** | **CRPS↓** ($10^{-2}$) (ensemble accuracy) | **AE↓** (motion direction) | **EPE↓** ($10^{-2}$) (motion displacement) | **FVD↓** ($10^{1}$) (spatiotemporal realism) | **WD↓** ($10^{-2}$) (distributional shift) | **twCRPS↓** ($10^{-2}$) (tail accuracy) |
|---|---|---|---|---|---|---|
| IMERG | 13.88 | 1.36 | 4.24 | 1.92 | 2.90 | 5.91 |
| Attention UNet | 9.96 | 1.43 | 4.33 | 4.52 | 5.19 | 4.35 |
| DDPM | 7.61 | 1.74 | 5.68 | 2.26 | 0.88 | **3.61** |
| ResDDPM | 7.89 | 1.59 | 5.17 | 1.81 | 1.28 | 3.77 |
| **HiDFiP (ours)** | **7.58** | **1.12** | **3.58** | **1.02** | **0.26** | 3.66 |

**Note:** Arrows indicate the direction of better performance. Rescaling factors are given in the column headers. Best values in Table S1 are shown in bold. For deterministic baselines (IMERG and Attention UNet), CRPS and twCRPS reduce to MAE and threshold-censored MAE, respectively.